\documentclass[aps,physrev,twocolumn,groupedaddress]{revtex4-2}

\usepackage[toc,page]{appendix}
\usepackage{amssymb,mathtools,physics,amsfonts}

\usepackage{subcaption}
\usepackage{float}

\usepackage{amsthm}
\usepackage{amsmath}
\usepackage{bbm, dsfont}
\usepackage{graphicx}
\usepackage[usenames,dvipsnames]{color}
\usepackage[colorlinks=true,citecolor=magenta,urlcolor=blue]{hyperref}
\usepackage[table]{xcolor}
\usepackage{colortbl}
\usepackage{color}
\usepackage{multirow}
\usepackage{hhline}
\usepackage{booktabs}
\usepackage[normalem]{ulem}
\usepackage{siunitx}
\usepackage{enumitem}
\usepackage{braket}
\usepackage{siunitx}
\usepackage{textcomp}
\usepackage{booktabs}

\usepackage[capitalise]{cleveref}

\DeclareSIUnit\permille{\text{\textperthousand}}

\newcommand{\quotes}[1]{``#1''}

\newcommand{\affilationVigoOne}{Vigo Quantum Communication Center, University of Vigo, Vigo E-36310, Spain}
\newcommand{\affilationVigoTwo}{Escuela de Ingeniería de Telecomunicación, Department of Signal Theory and Communications, University of Vigo, Vigo E-36310, Spain}
\newcommand{\affilationVigoThree}{AtlanTTic Research Center, University of Vigo, Vigo E-36310, Spain}

\newcommand{\decoyrefs}{\cite{hwang_quantum_2003,decoy,wang_beating_2005}}

\begin{document}

\title{Modeling and Characterization of Arbitrary Order Pulse Correlations for Quantum Key Distribution}
\author{Ainhoa Agulleiro}\email{aagulleiro@vqcc.uvigo.es}
	\affiliation{\affilationVigoOne}
	\affiliation{\affilationVigoTwo}
	\affiliation{\affilationVigoThree}
	
\author{Fadri Gr\"unenfelder}
	\affiliation{\affilationVigoOne}
	\affiliation{\affilationVigoTwo}
	\affiliation{\affilationVigoThree}

\author{Margarida Pereira}
        \affiliation{\affilationVigoOne}
	\affiliation{\affilationVigoTwo}
	\affiliation{\affilationVigoThree}

\author{Guillermo Currás-Lorenzo}
        \affiliation{\affilationVigoOne}
	\affiliation{\affilationVigoTwo}
	\affiliation{\affilationVigoThree}
 
\author{Hugo Zbinden}
	\affiliation{\affilationVigoOne}
	\affiliation{\affilationVigoTwo}
	\affiliation{\affilationVigoThree}

\author{Marcos Curty}
        \affiliation{\affilationVigoOne}
	\affiliation{\affilationVigoTwo}
	\affiliation{\affilationVigoThree}
	
\author{Davide Rusca}
	\affiliation{\affilationVigoOne}
	\affiliation{\affilationVigoTwo}
	\affiliation{\affilationVigoThree}

\date{\today}

\begin{abstract}
    In quantum key distribution (QKD) implementations, memory effects caused by the limited bandwidth of modulators and/or other active devices can leak information about previous setting choices. Security proofs addressing this imperfection require the characterization of pulse correlations, which, in principle, can be of an arbitrary order, even unbounded. Experimentally, this is very hard (if not impossible) to achieve. Here, we solve this pressing problem by introducing a simple linear model to explain pulse correlations. In so doing, we can derive upper bounds on the correlation strength of arbitrary order from the study of the step response of the system. Importantly, this is what is needed to ensure the security of QKD in the presence of pulse correlations of unbounded length. We experimentally characterize short-range correlations and apply the proposed method to account for long-range correlations to an infinite order.
\end{abstract}

\maketitle

\section{Introduction}\label{section: introduction}

Quantum cryptography has emerged as an attractive alternative to its classical counterpart due to its security being provided by the laws of quantum physics rather than relying on computational hardness. More concretely, quantum key distribution (QKD) has been shown to be information-theoretically secure when combined with the one-time pad protocol \cite{renner_security_2006}. Although most of its security proofs assume perfect devices \cite{shor_preskill,gottesman_proof_2003}, in practice, device flaws result in leakage of information that may be exploited by an eavesdropper (Eve) to learn the secret key without being detected. For this reason, a big effort has been put recently into trying to bridge the gap between theory and experiments \cite{hwang_quantum_2003,decoy,wang_beating_2005,GLLP,LT}.

In recent years, QKD has seen a fast development, with some state of the art demonstrations achieving secret key rates in the order of tens of Mbps \cite{grunenfelder2020performance,li_high-rate_2023,grunenfelder_fast_2023}. However, this comes at a cost: modulation devices have limited bandwidths and operating them at high repetition rates causes memory effects (also known as the patterning effect). This introduces setting-choice dependent (SCD) pulse correlations in the emitted states, which leak information about past setting choices. This poses a significant security threat.  On the practical side, some countermeasures have been proposed to mitigate this side-channel \cite{yoshino2018,lu_intensity_modulator_2021,kang_patterning-effect_2023,lu_intensity_2023}. On the theoretical side, various security proofs have been introduced to tackle correlations that can be present in the prepared states \cite{Zapatero_2021,Sixto_2022,RTpereira_quantum_2020,LTQCoin,imperfectPR,pereira2023,overcoming_intensity_fluctuations}. Even though these security proofs consider a finite and known maximum correlation length, $l_c$, they can in principle be extended to the more realistic case of unbounded correlations by combining them with the analysis of \cite{pereira2024quantum}, thus ensuring the security of QKD in the presence of this kind of source imperfection. Importantly, their application involves the use of parameters that have to be experimentally characterized.

After the first experimental report of intensity correlations presented in \cite{yoshino2018}, other works have focused on the characterization of this side-channel. Correlations between nearest-neighbor pulses in the encoding of the bits, basis and decoy intensity values were measured in \cite{grunenfelder2020performance} by directly observing the polarization and intensity of the states conditioned on the previous setting choice with a polarimeter and an optical oscilloscope, respectively. Trefilov \textit{et al.} characterized intensity correlations up to the sixth order and showed that higher-order correlations may, in some cases, be stronger than nearest-neighbors \cite{trefilov}. Xing \textit{et al.} proposed a real-time method for estimating intensity correlations from the detection statistics of single-photon detectors without the need of changing the experimental setup, and then applied this method to measure first-order correlations \cite{xing_characterization_2024}. These works highlight a fundamental limit that, to the best of our knowledge, affect all characterization methods used so far. Namely, they can only be applied up to a finite order, even though, in practice, correlations can be of arbitrary order (even unbounded). To guarantee security, long-range correlations must also be characterized. Moreover, characterizing correlations of very high order requires the generation of a large number of combinations of long trains of pulses. Indeed, the amount of data that needs to be acquired and processed scales exponentially with the considered $l_c$, which also poses a practical challenge. Furthermore, it is expected that correlations eventually become so small that they are indistinguishable from experimental noise.

In this work, we solve this pressing issue by proposing a new method to characterize SCD pulse correlations of arbitrary order. To do so, we model the bandwidth-limited devices within the transmitter as a linear time invariant (LTI) system acting as a low-pass filter. The linearity of the model simplifies the characterization because it allows the prediction of correlations of any order from the knowledge of only the step response of the LTI system. Moreover, the model is versatile in the sense that it can be chosen to accommodate the flaws introduced by any bandwidth-limited devices within the transmitter. Such a model also gives further insight into the imperfect state preparation, which may be exploited in other ways. For example, it can be used to distinguish correlations from state preparation flaws (SPFs). This distinction is crucial to achieve high-performance QKD because SPFs, contrary to correlations, barely have an impact on the secret-key rate (SKR) \cite{LTQCoin}. Lastly, we show that our model can be bounded by an exponential function. This allows us to estimate for the first time the rate at which the correlation strength decreases as the distance between the pulses increases. Importantly, this is what is needed to ensure the security of QKD in the presence of arbitrary-long correlations according to the analysis of \cite{pereira2024quantum}.  

Additionally, we present experimental results using a dual characterization method. For short-range correlations --- which for us means correlations up to the fourth order --- we measure the phase and intensity modulation pulses used for the encoding of the bit, basis and intensity levels necessary for the decoy-state method. Similarly to previous approaches, e.g. \cite{yoshino2018,grunenfelder2020performance,trefilov,xing_characterization_2024}, we generate all possible combinations of settings and compare the results to gauge the effect that changing past setting choices has on the actual signal. This allows us to estimate the strength of the short-range correlations. For long-range correlations, on the other hand, we use the data which is acquired for the characterization of short-range correlations to determine the parameters of the LTI system of our model. We then use these results to calculate the contribution of the unaccounted correlations up to infinite order.

The proposed model can be adapted to actual fast QKD setups. Its use makes the implementation of the existing security proofs tackling correlations more practical. More concretely, it provides a way to experimentally characterize the parameter that is needed for the analysis of \cite{pereira2024quantum}, thus taking a fundamental step further towards implementation security of QKD.

The structure of this paper is as follows. In \cref{section: theoretical framework}, we present the specific scenarios that we consider in this work and we introduce the metric that is used for quantifying SCD pulse correlations. In \cref{section: characterization of short-range correlations}, we explain how short-range correlations are experimentally characterized from the measurement of the modulation pulses. Also, we show the results of using this method for the scenarios considered. In \cref{section: characterization of long-range correlations}, we introduce our novel method for characterizing long-range pulse correlations by modeling our setup. We show that this model confirms the expected exponential decay of correlations with pulse separation as assumed in \cite{pereira2024quantum}, and we use the data acquired for the characterization of short-range correlations to provide the first ever estimation of long-range correlations. In \cref{section: SKR simulations}, we simulate the achievable SKR for a standard BB84 protocol with unbounded pulse correlations when using the proposed characterization method of SCD correlations. Lastly, the conclusions are summarized in \cref{section: conclusions}.

\section{Theoretical framework}\label{section: theoretical framework}

Let us consider an imperfect source with SCD pulse correlations, which cause the state of a certain round to depend not only on the setting choices made in that round but also on the previous ones. The device or devices responsible for the memory effects will affect the states in different ways depending on their role within the setup. For example, if the memory effects appear in a phase modulator (PM), whose role is to encode information in the relative phase between a reference and signal pulse, the prepared states will present bit-and-basis correlations. On the other hand, if the memory effects are present in the intensity modulator (IM) in charge of varying the intensities of weak coherent pulses (WCPs) in decoy-state QKD \decoyrefs, the states will have intensity correlations.

To apply security proofs accounting for SCD pulse correlations such as those introduced in \cite{Zapatero_2021,Sixto_2022,RTpereira_quantum_2020,LTQCoin,pereira2023,pereira2024quantum,overcoming_intensity_fluctuations}, a quantitative measure of their strength, denoted by $\epsilon_l$, is needed. Here, $l$ is the order of the correlation, i.e. the number of rounds that separates the pair of pulses under consideration. 
This parameter takes values $l=1, 2, ..., l_{\rm c}$, with $l_{\rm c}$ being the maximum correlation length, as already introduced. To quantify $\epsilon_l$, we consider the maximum variation in the $N^{\rm th}$ state when changing the $(N-l)^{\rm th}$ setting, $N$ being the total number of rounds. This can be related to a lower bound on the fidelity \cite{RTpereira_quantum_2020,LTQCoin,pereira2024quantum}, that is,
\begin{widetext}
\begin{equation}\label{definition epsilon_l}
    F\left(\rho_{j_N|j_{N-1}, ..., j_{N-l}, ..., j_1}
    , \rho_{j_N|j_{N-1}, ..., \Tilde{j}_{N-l}, ..., j_1}\right) \geq 1 - \epsilon_l,
\end{equation}
\end{widetext}
where $j_k$ for $k \in \{1, ...,N\}$ is the setting selected in the $k^{\rm th}$ round, $F(\rho, \sigma)$ is the fidelity between $\rho$ and $\sigma$, and $\rho_{j_N|j_{N-1}, ..., j_1}$ is the state in the $N^{\rm th}$ round conditioned on the settings selected in the previous rounds. Importantly, we consider the $N^{\rm th}$ pulse because it is the one affected by the largest number of previous setting choices. Thus, the correlation strengths are given by
\begin{widetext}
\begin{equation}\label{definition of correlation strength from fidelity}
    \epsilon_l := \max_{\mathbf{j}_N,\Tilde{\mathbf{j}}_N}\left\{ 1 -  F\left(\rho_{j_N|j_{N-1}, ..., j_{N-l}, ...,j_1}
    , \rho_{j_N|j_{N-1}, ..., \Tilde{j}_{N-l}, ...,j_1}\right) \right\}.
\end{equation}
\end{widetext}
Here, the maximum is taken over all possible combinations of settings $\mathbf{j}_N:=j_1,...,j_{N-l},...,j_N$ and  $\Tilde{\mathbf{j}}_N:=j_1,...,\Tilde{j}_{N-l},...,j_N$, where the latter differs from the former only in the settings selected in the $(N-l)^{\rm th}$ round. In this work, we shall use the superscript $\mu$ ($\phi$) to indicate results for intensity (phase) correlations. For instance, $\epsilon_l^{\mu}$ ($\epsilon_l^{\phi}$) denotes the strength of intensity (phase) correlations of order $l$.

\begin{figure*}
    \centering
    \includegraphics[width=0.6\linewidth]{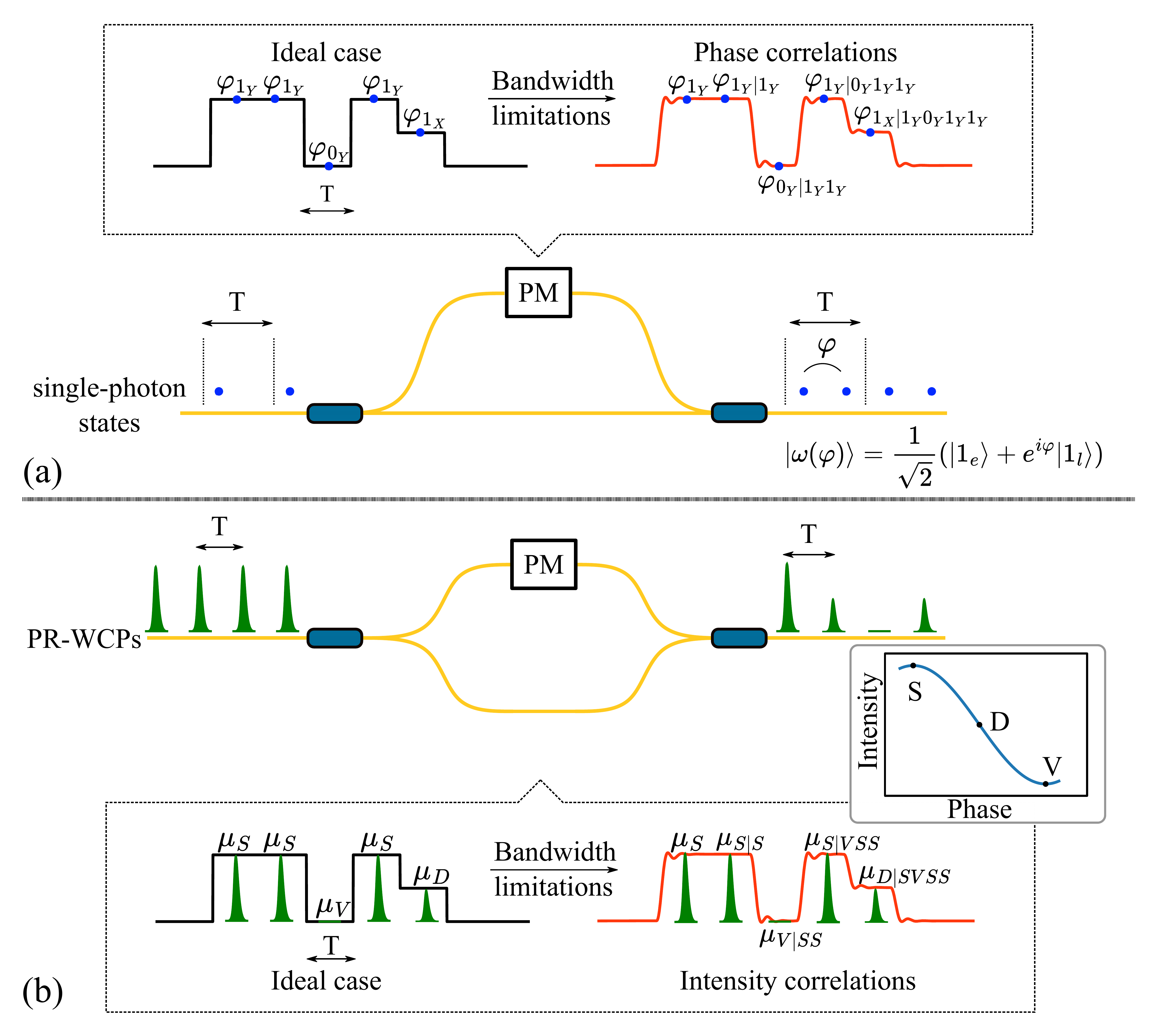}
    \caption{PM: phase modulator, PR-WCPs: phase-randomized weak coherent pulses, S: signal, D: decoy, V: vacuum. (a) Schematic representation of the implementation of a phase-encoding setup with an unbalanced Mach-Zehnder interferometer (MZI). The difference in the length of the arms splits the state into two temporal modes. The PM introduces a relative phase between them. The upper part of the figure shows how bandwidth limitations of the PM cause phase correlations. (b) Schematic representation of the intensity modulation of PR-WCPs through the action of an IM consisting of a balanced MZI with a PM in one of its arms. As shown in the small inset plot, the phase of the PM controls the result of the interference, thus changing the intensity of the state. Therefore, bandwidth limitations of the PM cause intensity correlations, which is shown in the lower part of the figure.}
    \label{fig:balanced and unbalanced MZI}
\end{figure*}

In the two following subsections, we introduce the concrete scenarios that are considered in this work, which are schematically represented in \cref{fig:balanced and unbalanced MZI}.

\subsection*{Phase correlations}

Firstly, we consider correlations in the encoding of the bit-and-basis, which, in this case, is assumed for concreteness to be done via phase modulation. A similar analysis could be done for instance for polarization encoding. For simplicity, we consider single-photon states. Note, however, that the characterization method we describe in this section can also be applied to the single-photon components of phase-randomized (PR) WCPs. 

In particular, let us consider that, as shown in \cref{fig:balanced and unbalanced MZI} (a), Alice uses an unbalanced Mach-Zehnder interferometer (MZI) to encode information in the relative phase between two temporal modes of the single-photon states. The MZI splits the states into two temporal modes denoted by early and late. In the long arm of the MZI, a PM is placed, which introduces a relative phase $\varphi$. In general, these phases are selected using a series of control electric pulses. Ideally, each of these electric pulses should be square and its height should be determined only by the setting choice corresponding to that round. Realistically, bandwidth limitations of the PM and/or the devices driving it cause their shape to differ from square pulses and introduce SCD correlations. Moreover, they provoke a time dependence of the phase inside the time-bins, which causes information to be encoded in time as reported in \cite{gnanapandithan_hidden_2025}. Our focus is solely on SCD correlations, so for simplicity in this work we disregard the latter side-channel by considering that the temporal distribution of the states is sufficiently narrow that it can be approximated to a Dirac delta. With all these considerations, one can write the states prepared by Alice as
\begin{equation}\label{phase-encoded qubit}
    \ket{\psi\left(\varphi(t_0)\right)} = \frac{1}{\sqrt{2}}\left(\ket{1_{\rm e}} + e^{i\varphi(t_0)}\ket{1_{\rm l}}\right),
\end{equation}
where the subscripts $e$ and $l$ refer to the early and late time bins, respectively. They are taken as the eigenstates of the computational basis. The parameter $t_0$ is the position of the emitted signal pulses within the modulation pulses. It is defined such that $t_0=0$ refers to the starting edge of the modulation pulse and $t_0=T$ to the end, $T$ being both the width of the modulation pulse and the separation between the emitted signals.

To quantify the strength of the correlations, we consider a scenario where Alice chooses the setting $j_N$ for the $N^{\rm th}$ pulse, which corresponds to the ideal phase $\phi_{j_N}$. Because of correlations, the actual phase value depends on the previous setting choices $j_{N-1}$,$j_{N-2}$,...,$j_{N-l}$,...,$j_1$. We denote this actual phase by $\varphi_{j_N|\mathbf{j}_{N-1}}(t_0) := \varphi_{j_N|j_{N-1}, ..., j_{N-l}, ..., j_1}(t_0)$. Then, we consider the phase when the exact same string of settings is used for the $N$ rounds except for the $(N-l)^{\rm th}$ round, $\varphi_{j_N|\mathbf{\Tilde{j}}_{N-1}}(t_0):=\varphi_{j_N|j_{N-1}, ..., \Tilde{j}_{N-l}, ..., j_1}(t_0)$. From \cref{definition of correlation strength from fidelity}, it follows that the correlation strength is given by 
\begin{widetext}
\begin{equation}\label{eq: experimental epsilon_l phi 2}
    \epsilon_l^{\phi}(t_0) := \max_{\mathbf{j}_N, \mathbf{\Tilde{j}}_N} \left\{ 1 - \cos^2 \left(\frac{\varphi_{j_N|\mathbf{\Tilde{j}}_{N-1}}(t_0) - \varphi_{j_N|\mathbf{j}_{N-1}}(t_0)}{2}\right)\right\}. 
\end{equation}
\end{widetext}
The explicit calculation of the fidelity used in \cref{eq: experimental epsilon_l phi 2} is shown in \cref{eq:explicit fidelity calculation  (phase)} in \cref{appendix: imperfect step response}.

\subsection*{Intensity correlations}

Let us now consider that Alice generates PR-WCPs to implement the decoy-state method. To this end, she changes their intensities with an IM that consists of a balanced MZI with a PM in one of its arms as shown in \cref{fig:balanced and unbalanced MZI} (b). The PM introduces a relative phase, which controls the result of the interference, thus modulating the intensity of the output pulses. Again, bandwidth limitations in the PM and/or the electrical signal driving it cause the phase signal to present memory effects. After interference, this translates into intensity correlations. We consider two WCPs with mean photon numbers that, in analogy to the phase scenario, we denote by $\mu_{j_N|\mathbf{j}_{N-1}}(t_0) := \mu_{j_N|j_{N-1}, ..., j_{N-l}, ..., j_1}(t_0)$ and $\mu_{j_N|\mathbf{\Tilde{j}}_{N-1}}(t_0) := \mu_{j_N|j_{N-1}, ..., \Tilde{j}_{N-l}, ..., j_1}(t_0)$. These values are the result of modulating the $N^{\rm th}$ pulse when the same string of \textit{N }settings is used except for the $(N-l)^{\rm th}$ round. For simplicity, here we assume perfect phase randomization of the WCPs so that they are diagonal in the Fock basis, and that the intensity correlations do not change the Poissonian distribution of the photon number conditioned on the actual intensity level \cite{Zapatero_2021, Sixto_2022}. Therefore, in this scenario the fidelity is given by the  classical fidelity of two Poissonian distributions. Thus, the correlation strengths defined in \cref{definition of correlation strength from fidelity} may be expressed as
\begin{widetext}
\begin{equation}\label{eq: experimental intensity correlations}
    \epsilon_l^{\mu}(t_0) := \max_{\mathbf{j}, \mathbf{\Tilde{j}}}\left\{1 - \left(  \sum_{n=0}^{\infty} \frac{\sqrt{e^{-\left[\mu_{j_N|\mathbf{j}_{N-1}}(t_0)+\mu_{j_N|\mathbf{\Tilde{j}}_{N-1}}(t_0)\right]}\left[\mu_{j_N|\mathbf{j}_{N-1}}(t_0)\mu_{j_N|\mathbf{\Tilde{j}}_{N-1}}(t_0)\right]^n}}{n!} \right)^2 \right\}.
\end{equation}
\end{widetext}
The explicit calculation of the fidelity used for \cref{eq: experimental intensity correlations} is shown in \cref{fidelity of two Poissonians,eq: Poissonian distribution} in \cref{appendix: imperfect step response}.

\section{Characterization of short-range correlations}\label{section: characterization of short-range correlations}

\begin{figure*}
    \centering
    \includegraphics[width=0.75\linewidth]{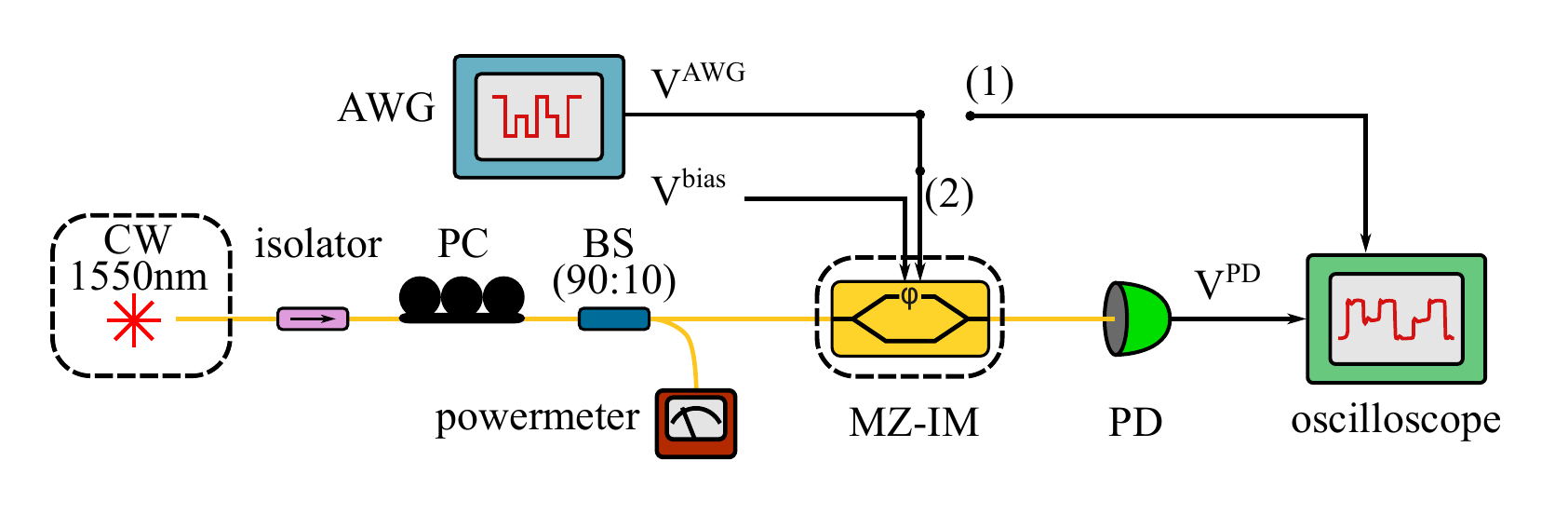}
    \caption{Experimental setup. The dashed lines indicate a thermally controlled box. PC: polarization controller, AWG: arbitrary waveform generator, MZ-IM: Mach-Zehnder intensity modulator, BS: beam splitter, PD: linear photodetector. The output of the AWG can be directly connected to (1) the oscilloscope to measure the voltage from the AWG, $V^{\rm AWG}(t)$, or to (2) the IM in order to modulate the output of the CW laser. In the latter case, the oscilloscope measures the voltage from the PD, $V^{\rm PD}(t)$.}
    \label{fig:experimental setup}
\end{figure*}

In this section, we use a purely experimental approach to characterize the correlations up to fourth order. To do so, we measure the imperfect modulation pulses that introduce SCD pulse correlations. A schematic drawing of the experimental setup is shown in \cref{fig:experimental setup}. A 1550nm laser is operated in continuous wave (CW). An isolator is added afterwards to reduce backscattering into the cavity of the laser. The intensity of the light generated by the laser is modulated to the signal (S), decoy (D) and vacuum (V) intensity levels by an IM with a bandwidth of 10GHz (MXER-LN-10). A polarization controller is placed between the laser and the IM so that the polarization state of the signals entering the IM is optimal to maximize the extinction ratio.

The IM consists of a balanced MZI with a PM in one of its arms. A DC bias voltage controls its working point by providing a constant phase shift. An arbitrary waveform generator (AWG) of bandwidth 150MHz (-3dB) (Tektronix AFG31102) controls the phase and, consequently, intensity modulation through an RF voltage $V^{\rm AWG}(t)$. This is the device with smallest bandwidth within the setup, so we restrict ourselves to a repetition rate of 50MHz.  The nominal half-wave voltage $V_{\pi}$ is 6V. This quantity relates the phase $\varphi$ introduced by the PM to the applied voltage with the equation given by
\begin{equation}\label{eq: action of PM}
    \varphi(t_0) = \frac{V^{\rm AWG}(t_0)}{V_{\pi}}\pi + \pi.
\end{equation}
Here, the constant $\pi$ offset comes from the action of the bias voltage.

The optical output of the IM is injected into a linear photodetector (PD) with a bandwidth of 23.9GHz (Alphalas UPD-15-IR2-FC), which provides a voltage $V^{\rm PD}(t)$. Both $V^{\rm AWG}(t)$ and $V^{\rm PD}(t)$ are measured with a digital oscilloscope with bandwidth of 4GHz (Tektronix MSO 64B). The temperature of the laser and the IM are monitored and kept at constant values of 20ºC and 21ºC ($\pm$5mK), respectively. 

In this setup, the memory effects causing SCD correlations can, in principle, be generated in the AWG and/or the PM inside the IM. However, given that their bandwidths differ in two orders of magnitude and that the repetition rate is very low compared to the bandwidth of the PM, we consider the correlations to originate in the AWG only. 

The AWG is set to generate square pulses at a repetition rate of 50MHz and a duty cycle of 100$\%$ (i.e., pulse width $T$ of 20ns), and heights chosen from $\{- V_{\pi}/2, 0,  V_{\pi}/2\}$. In the phase-encoding scenario, this corresponds to preparing the states $\ket{0_Y}$, $\ket{1_X}$ and $\ket{1_Y}$, respectively. In the decoy-state protocol scenario, this corresponds to generating the signal, decoy and vacuum states, respectively.

All 243 possible sequences of 5 pulses are generated with the aim of quantifying the effect that the first four pulses have on the fifth one. To do so, we first determine the phase and intensity modulation pulses. The phase signal can be directly calculated with \cref{eq: action of PM} from the measurement of the voltage $V^{\rm AWG}(t)$. Intensity modulation pulses are directly measured with $V^{\rm PD}(t)$. In \cref{Appendix: intensity correlations from AWG}, an alternative method is used in which the intensity correlations are instead calculated from $V^{\rm AWG}(t)$ by using the parameters of the IM and calculating the result of interference.

\subsection{Phase correlations}\label{section: short-range phase correlations}

The phase modulation pulses are directly measured by taking oscillograms of $V^{\rm AWG}(t)$ and using \cref{eq: action of PM}. The data corresponding to all the 243 combinations is used to calculate the correlation strengths given by \cref{eq: experimental epsilon_l phi 2}. 
The value of $\epsilon_l^{\phi}(t_0)$ varies greatly with the chosen $t_0$. For instance, this parameter increases significantly when we get too close to the edges of the modulation pulse. It is, however, more practical to focus on one good realistic value of $t_0$, which we shall denote by $t_0^b$ for best-case. For reference, we also take another point denoted by $t_0^w$ for worst-case. To choose $t_0^b$, we  optimize over the quantity $\epsilon_{\rm total}^{\phi}(t_0)=\epsilon_{\rm qubit}^{\phi}(t_0)+\epsilon_{\rm correl}^{\phi}(t_0)$ defined below, since this is the parameter needed to apply the existing security analyses \cite{LTQCoin,RTpereira_quantum_2020,pereira2023}.  After finding the value of $t_0^b$ that minimizes $\epsilon_{\rm total}^{\phi}(t_0)$ in a region that is not too close to the edges, we choose $t_0^w$ as the point inside an interval of 4 ns around $t_0^b$ that maximizes $\epsilon_{\rm total}^{\phi}(t_0)$. This interval takes into account the possible jitter or misalignment of the states. In an actual QKD implementation, this parameter should also be characterized.

Precisely, the parameter $\epsilon_{\rm qubit}^{\phi}(t_0)$ quantifies how different the emitted state is from a certain qubit state, whereas $\epsilon_{\rm correl}^{\phi}(t_0)$ quantifies the amount of information leaked through the following $l_{\rm c}$ pulses. When there are no additional side-channels, $\epsilon_{\rm qubit}^{\phi}(t_0)$ and $\epsilon_{\rm correl}^{\phi}(t_0)$ can be expressed in terms of the correlation strengths $\epsilon_l^{\phi}(t_0)$ as 
\begin{subequations}\label{eq:epsilon_qubit epsilon_correl}
    \begin{equation}\label{eq:epsilon_qubit epsilon_correl 1}
        \epsilon_{\rm correl}^{\phi}(t_0) = \sum_{l=1}^{l_{\rm c}} \epsilon_l^{\phi}(t_0) ~~~\text{and}~~~  
    \end{equation}
    \begin{equation}\label{eq:epsilon_qubit epsilon_correl 2}
        \epsilon_{\rm qubit}^{\phi}(t_0) = \left(\sum_{l=1}^{l_{\rm c}}\sqrt{\epsilon_l^{\phi}(t_0)}\right)^2, 
    \end{equation}
\end{subequations}
with $l_{\rm c}=4$ in our case because we only consider combinations of 5 pulses. For further details on these parameters and how they are calculated, see \cref{appendix: calculation of epsilon}.

\cref{fig: results phase correlations}(a) shows the value of $\epsilon_{\rm total}^{\phi}(t_0)$ as a function of the point of alignment $t_0$. The first edge of the modulation pulse contains the most information about the previous setting choice, since the system has not had enough time to respond to the change of setting. Therefore, $\epsilon_{\rm total}^{\phi}(t_0)$ is very high in the first 3ns but quickly decreases inside the rest of the time bin. Interestingly, there is a noticeable increase between $t_0 \approx 5ns$ and $t_0\approx 10ns$. This is due to the ringing artifact, small relaxation oscillations that cause the signal to oscillate around the ideal value. These oscillations can be observed in the plot of $V^{\rm AWG}(t)$ shown in \cref{fig: fit} (a).

Following the optimization strategy described previously, we choose $t_0^b= 17.96 \rm ns$ 
and $t_0^w=16.0 \rm ns$ 
(marked as triangles pointing upwards and downwards in \cref{fig: results phase correlations} (a), respectively). The correlation strengths for these two values of $t_0$ are shown in \cref{fig: results phase correlations}(b). As expected, the correlations decrease with the pulse separation.

\begin{figure}[h!]
    \centering
    \begin{subfigure}[h]{0.45\textwidth}
        \includegraphics[width=\linewidth]{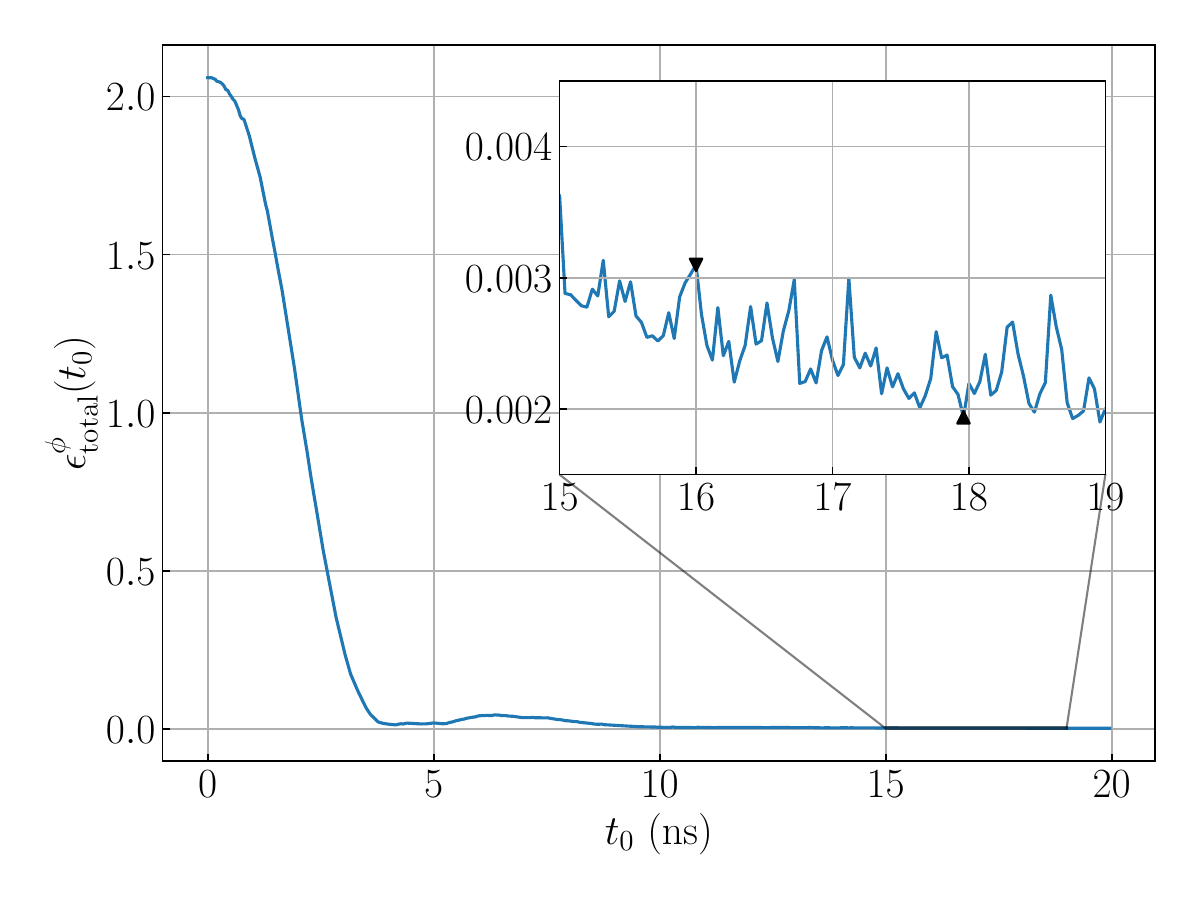}
        \caption{}
    \end{subfigure}
    \begin{subfigure}[h]{0.45\textwidth}
        \includegraphics[width=\linewidth]{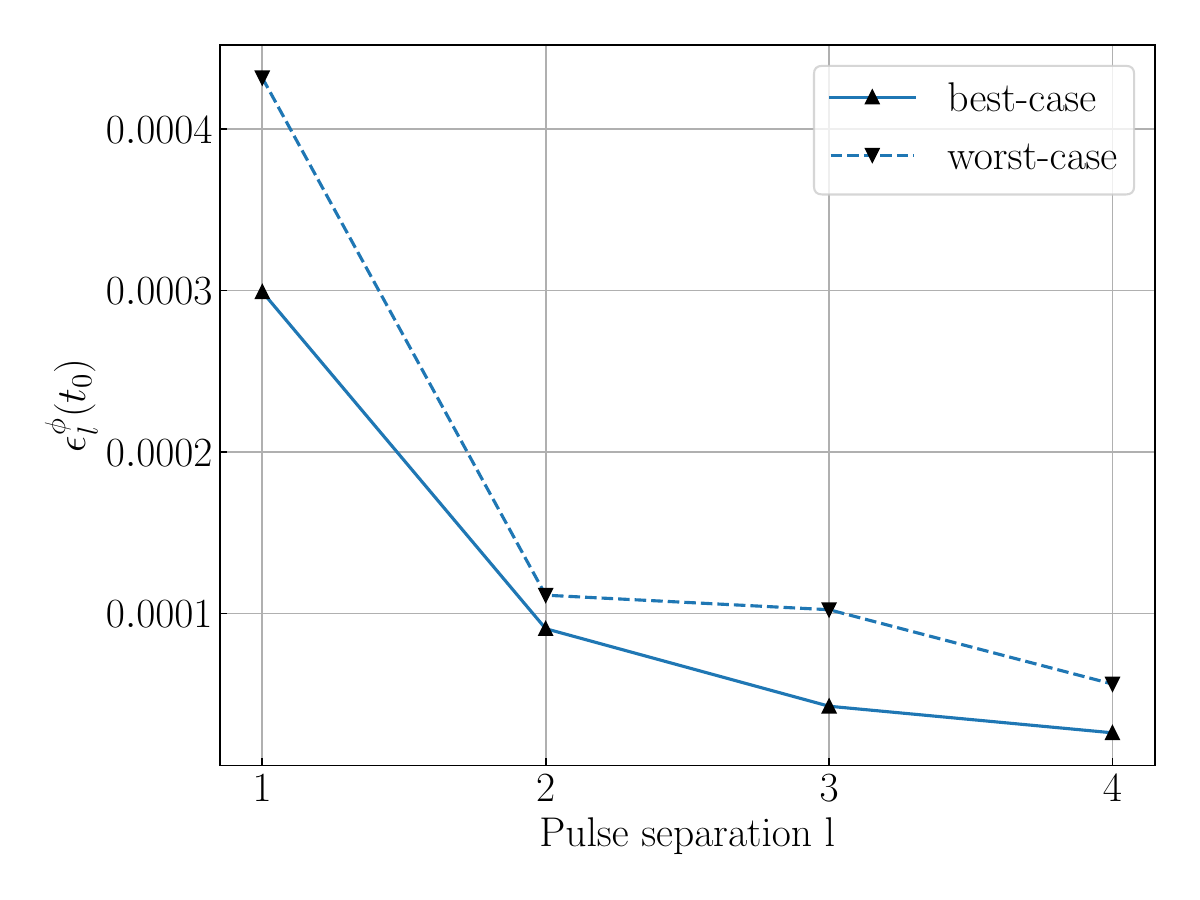}
        \caption{}
    \end{subfigure}
    \caption{(a) $\epsilon_{\rm total}^{\phi}(t_0)$ as a function of the point of alignment $t_0$. In the zoomed in region, the triangle pointing upwards marks the minimum, which occurs at $t_0^b=17.96 \rm ns$, 
    whereas the triangle pointing downwards shows the maximum in a 4 ns interval centered around the minimum, which occurs at $t_0^w=16.0\rm ns$. 
    (b) Phase correlation strengths $\epsilon_l^{\phi}(t_0)$ estimated for $t_0=t_0^{b}$ (solid line and triangles pointing upwards) and $t_0=t_0^{w}$ (dashed line and triangles pointing downwards) against the pulse separation between signals, $l=1, ..., 4$.}
    \label{fig: results phase correlations}
\end{figure}

\subsection{Intensity correlations}\label{section: short-range intensity correlations}

\begin{figure}[h]
    \centering
    \includegraphics[width=\linewidth]{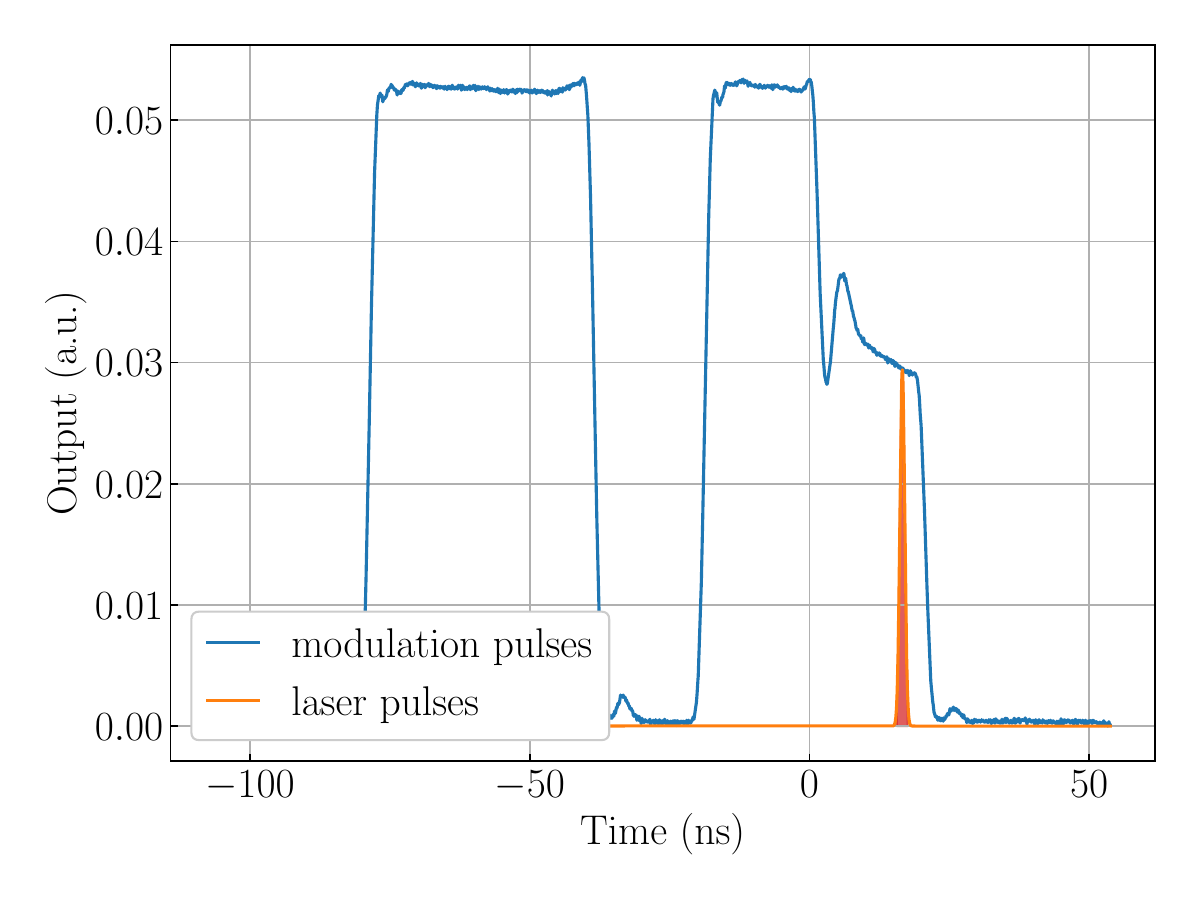}
    \caption{Plot showing how the mean photon number is calculated from $V^{\rm PD}(t)$ using \cref{eq: mu out}. The red area under the laser pulse curve is the integration area used for this calculation. The origin of the time axis is chosen at $t_0=0$.}
    \label{fig:laser pulses}
\end{figure}

The intensity modulation pulses are directly measured by taking oscillograms of $V^{\rm PD}(t)$. Again, data is acquired for all the 243 possible combinations of setting choices to calculate $\epsilon_l^{\mu}(t_0)$ using \cref{eq: experimental intensity correlations}. For this, we need to estimate the mean photon number $\mu$ of the modulated laser pulses. As shown in \cref{fig:laser pulses}, this is done by integrating them over time, that is,
\begin{equation}\label{eq: mu out}
    \mu(t_0) \propto \int_{t_0-\Delta t/2}^{t_0 +\Delta t /2} 
    f(t-t_0)V^{\rm PD}(t)dt,
\end{equation}
where $\Delta t$ is the integration width and $f(t-t_0)$ is the temporal profile of the unmodulated laser pulses. For concreteness, we shall consider that they follow a Gaussian profile, i.e.,
\begin{equation}\label{eq: gaussian pulse}
    f(t-t_0) = \dfrac{1}{\sigma\sqrt{2\pi}}e^{-\dfrac{(t-t_0)^2}{2\sigma^2}},
\end{equation}
where $\sigma$ is related to the full width at half maximum (FWHM) of the pulse through the equation $\rm FWHM=2\sqrt{2\ln 2}\sigma$. We remark, however, that the same analysis could be done for any other arbitrary shape. 

The proportionality factor in \cref{eq: mu out} is chosen so that the mean photon number of the signal states is 0.3. We choose this value according to \cite{grunenfelder2020performance}. In the spirit of preserving as well the order of the ratio between the FWHM and the duration $T$ between pulses of \cite{grunenfelder2020performance}, the FWHM is set to 1ns. The pulses are integrated over $\Delta t=2 \rm FWHM$. The obtained values of $\mu(t_0)$ are used in \cref{eq: experimental intensity correlations} to calculate $\epsilon_l^{\mu}(t_0)$ for different values of $t_0$.

\begin{figure}[h!]
    \centering
    \begin{subfigure}[h]{0.45\textwidth}
        \includegraphics[width=\linewidth]{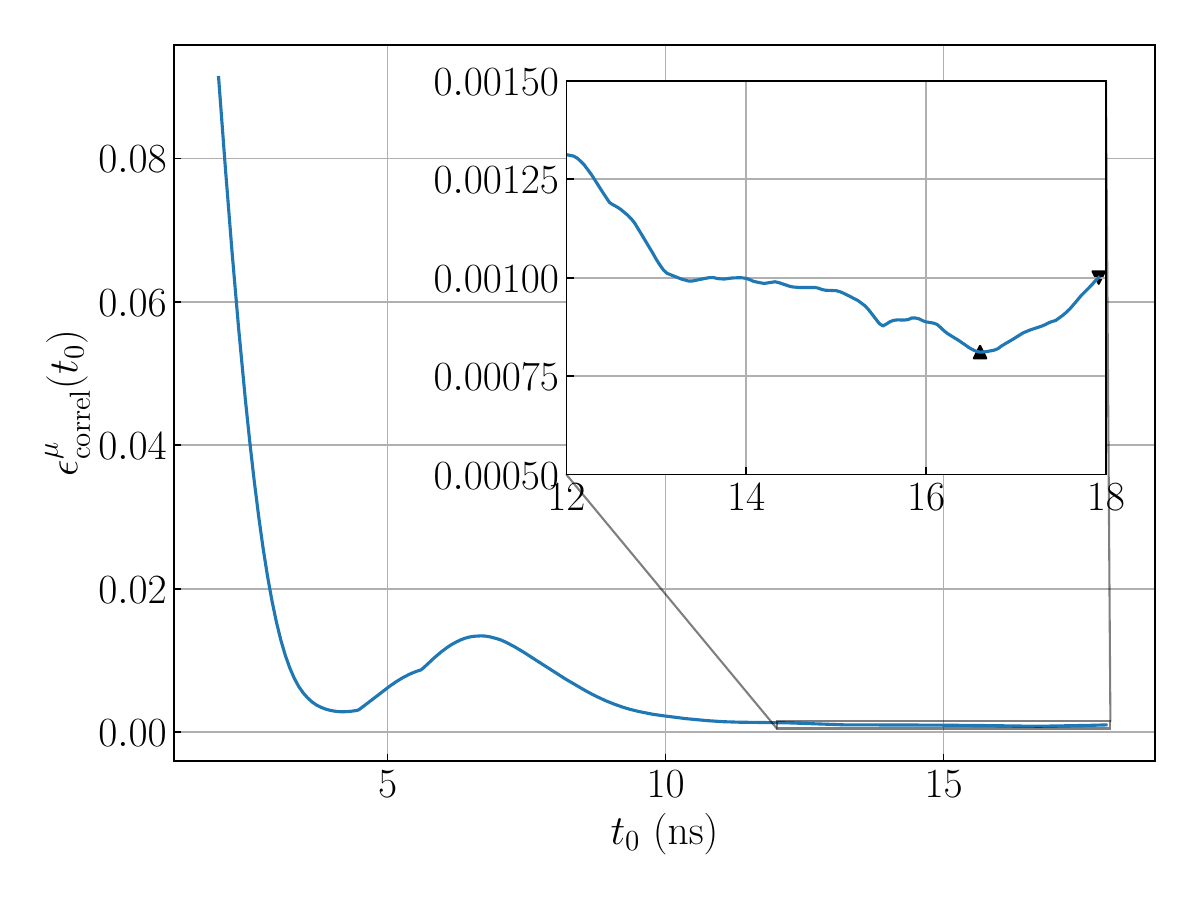}
        \caption{\textcolor{red}{}}
    \end{subfigure}
    \begin{subfigure}[h]{0.45\textwidth}
        \includegraphics[width=\linewidth]{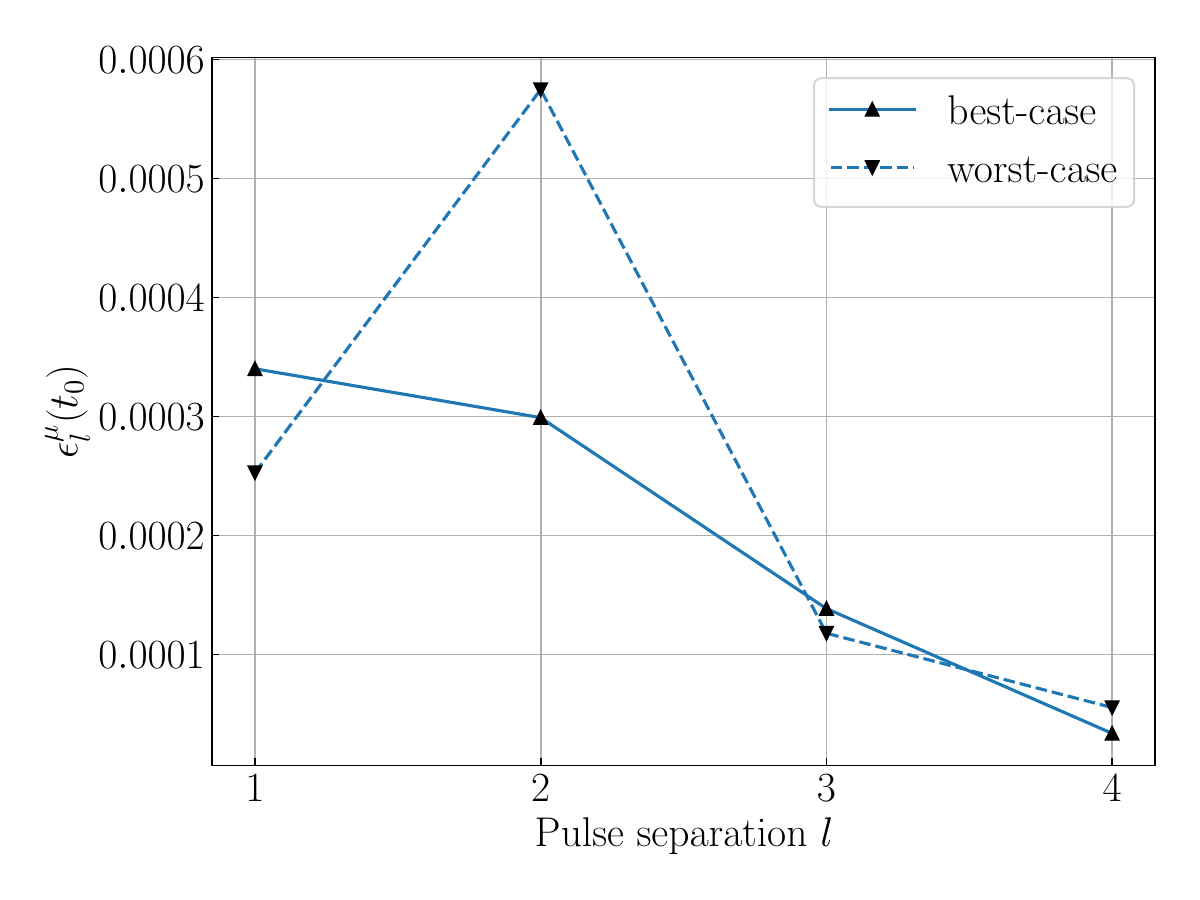}
        \caption{}
    \end{subfigure}
    \caption{(a) $\epsilon_{\rm correl}^{\mu}(t_0)$ against the point of alignment $t_0$ inside the modulation pulse calculated with data from the PD. According to these results, we choose $t_0^b=16.6\rm ns$ and $t_0^w=17.92\rm ns$. 
    (b) Maximum intensity correlation strength estimated from \cref{eq: experimental intensity correlations} for $t_0^{b}$ and $t_0^{w}$ as a function of the distance between signals $l=1, ..., 4$.}
    \label{fig: deciding t0 intensity experimental}
\end{figure}

As in the phase correlations scenario, we need a way to choose a point of alignment. To this end, we optimize over a parameter that we denote by $\epsilon_{\rm correl}^{\mu}(t_0)$. This parameter is defined in the same way as $\epsilon_{\rm correl}^{\phi}(t_0)$ in \cref{eq:epsilon_qubit epsilon_correl 1} but changing $\epsilon_l^{\phi}(t_0)$ with $\epsilon_l^{\mu}(t_0)$. Again, we use this parameter because it is needed to apply the existing security analyses for intensity correlations \cite{Zapatero_2021,Sixto_2022}. The results of calculating $\epsilon_{\rm correl}^{\mu}(t_0)$ are shown in \cref{fig: deciding t0 intensity experimental} (a). The corresponding chosen points of alignment are $t_0^b=16.6\rm ns$  
and $t_0^w=17.92\rm ns$. 
In this figure, we observe that the first 4ns still contain much of the information of the previous setting choices and that the effect of the ringing artifact is much more pronounced than in the phase scenario. 

The correlation strengths are displayed in \cref{fig: deciding t0 intensity experimental} (b) for the chosen $t_0^{b}$ and $t_0^{w}$. In the latter, we observe an increase for $l=2$. This might seem counterintuitive, but similar results were reported in \cite{trefilov}. The cause of this seems to be the ringing artifact. Interestingly, in contrast to the phase scenario, the intensity correlations vary greatly depending on the setting choice of the emitted pulse. This effect and its cause are explained in \cref{appendix: setting dependence}. 

Note that, as shown in \cref{Appendix: intensity correlations from AWG}, $\mu(t_0)$ could also be directly estimated from $V^{\rm AWG}(t)$. The correlation strengths estimated from the PD are higher. This is due to the fact that there are additional effects influencing the measurement results. In particular, the bias drift from the IM introduces more imperfections that influence the correlation strength, and, moreover, the PD introduces additional noise.

\section{Characterization of long-range correlations}\label{section: characterization of long-range correlations}

Due to physical limitations, so far we have only considered correlations up to fourth order. Nonetheless, as discussed before, correlations may exist across pulses that are very distant from one another. 

The security of QKD in the presence of unbounded pulse correlations can be ensured by applying the analysis of \cite{pereira2024quantum}. This is achieved by considering an \textit{effective} maximum correlation length $l_{\rm e}$, which should be chosen such that the residual correlations between pulses separated by more than $l_{\rm e}$ rounds are essentially negligible. In this context, negligible means that the actual scenario is indistinguishable from one in which these residual correlations are exactly zero, except with a very small failure probability $d$. Then, as shown in \cite{pereira2024quantum}, one can simply apply the existing analyses \cite{RTpereira_quantum_2020,mizutaniFinitekeySecurity2023,pereira2023,LTQCoin} as if the true maximum correlation length was indeed $l_{\rm e}$, and increase the security parameter by $2d$ to account for the residual correlations beyond this limit.

More concretely, the authors of \cite{pereira2024quantum} show that if one can find an upper bound $\overline{\epsilon}_l$ on the correlation strength $\epsilon_l$ that decays exponentially with the pulse separation $l$, that is,
\begin{equation}\label{eq: epsilon exponential}
    \epsilon_l \leq \overline{\epsilon}_l = \overline{\epsilon}_1 e^{-C(l-1)}
\end{equation}
for a certain $C>0$, then the effective maximum correlation length $l_{\rm e}$, takes the following form:
\begin{equation}\label{eq: correlation length}
    l_{\rm e} = \frac{1}{C}  \ln \left[\frac{N\overline{\epsilon}_1}{d^2 (1-\sqrt{e^{-C}})^2}\right],
\end{equation}
where $N$ is the total number of emitted signals. Importantly, to apply these results the validity of \cref{eq: epsilon exponential} has to be experimentally verified and the parameters $\overline{\epsilon}_1$ and $C$ must be experimentally characterized.

Therefore, it is necessary to introduce a way to characterize long-range correlations. In this section we propose a method to upper bound the strength of long-range correlations from the study of the step response of the system. To do so, we rely on the assumption that it is possible to model devices displaying memory effects as LTI systems. This is the case for a PM since it uses the Pockels effect in order to modulate the phase, which is a linear process with respect to the applied electric field. In other cases, the justification might not be so obvious, but we will assume that if the device has non-linearities, they are still dominated by its linear behavior. Because an IM --- consisting of a balanced MZI with a PM in one of its arms --- is inherently non-linear, we treat it differently by separating the linear and non-linear parts of its operation. Precisely, we shall consider that intensity correlations are caused in the PM --- which can still be modeled as an LTI system --- while interference (the origin of the non-linearities) is perfect. Thus, intensity correlations can still be estimated by first calculating the phase correlations with the model and then considering how these impact the result of interference.

\subsection{Exponential upper bounds for the correlation strengths}\label{section: exponential upper bounds for correlations}

\begin{figure}[h]
    \centering
    \includegraphics[width=0.8\linewidth]{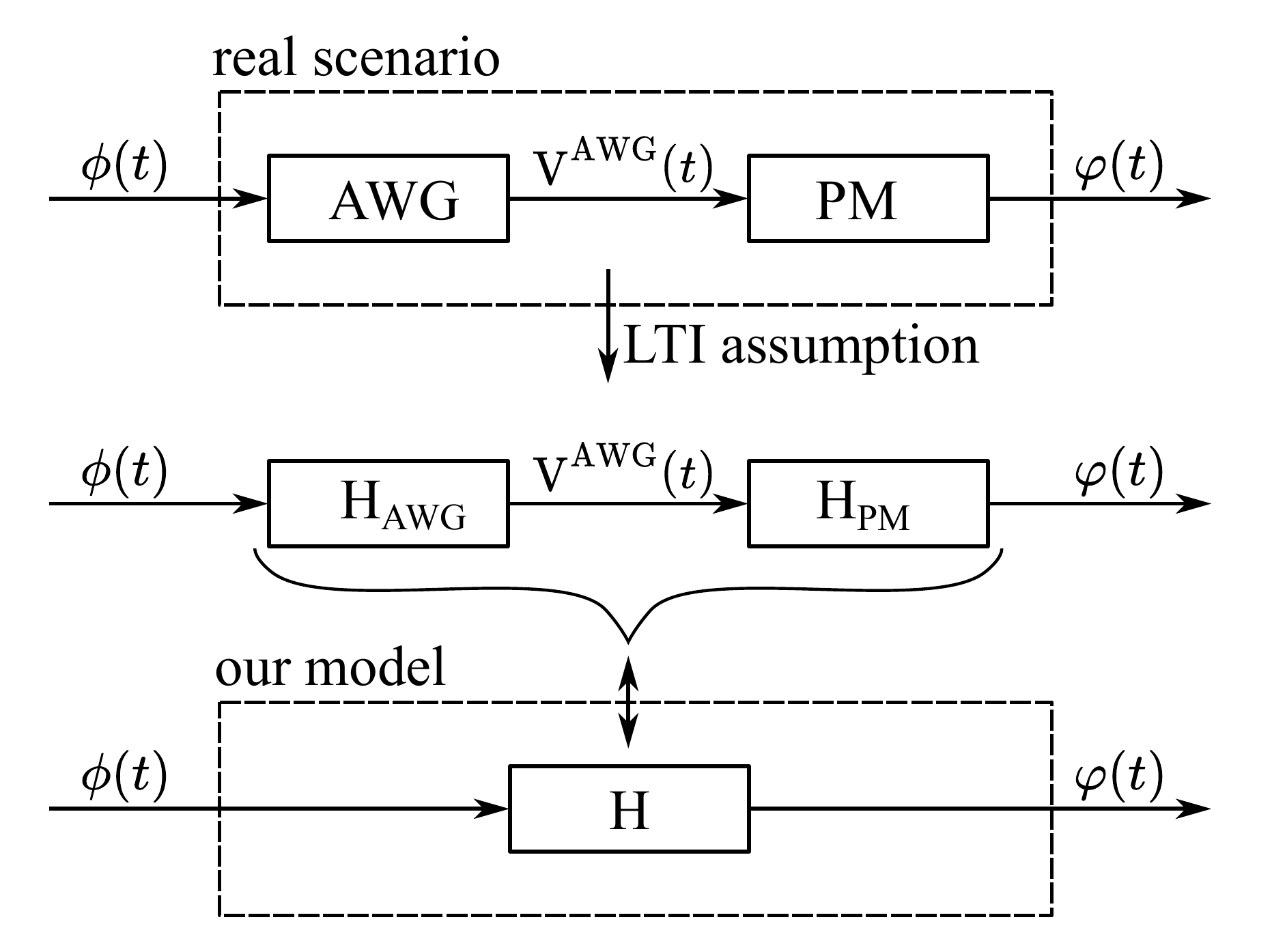}
    \caption{AWG: arbitrary waveform generator, PM: phase modulator. The variables $\phi(t)$ and $\varphi(t)$ are the input and output of the system, respectively. The former represents the ideal signal, while the latter represents the real signal with correlations. In the real scenario, $\phi(t)$ is first modified by the AWG, which provides a voltage $V^{\rm AWG}(t)$, and then by a PM, yielding the final real phase $\varphi(t)$. The LTI assumption allows us to model the behavior of these devices with their transfer functions, which we denote by $H_{\rm AWG}$ and $H_{\rm PM}$, respectively. Finally, for our model, we consider the equivalent scenario where all the bandwidth limited devices are encompassed in one LTI system with transfer function $H$. More generally, $H$ could accommodate an arbitrary number of devices. Thus, it is not crucial to know from which device or devices the imperfections come from as long as these imperfections are captured by $H$.}
    \label{fig: schematic of model}
\end{figure}

The transfer function $H$ of an LTI system models its response to any given input. In a bandwidth-limited device, its imperfect signal can thus be explained by transforming the ideal signal with the transfer function of the system. In a QKD experiment, the limitations might come from the AWG or the PM, inter alia. Indeed, in our experiment the voltage signal coming from the AWG is already imperfect and can explain the presence of correlations. Had we used a repetition rate in the order of GHz and a AWG with higher bandwidth, the PM would be expected to further modify the phase signal. However, for the sake of simplicity we consider a black box model (which encompasses the AWG, PM and any other bandwidth-limited devices) where the input is an ideal signal and the output are the correlated phase modulation pulses (see \cref{fig: schematic of model}).

When the input in consideration is the Heaviside step function $\theta(t)$, the output of the system is given by the so-called step response, which we denote by $\vartheta(t)$. Because of the linearity of the system, one can explain how the correlations arise in the phase modulation pulses by considering the difference between the actual step response and the ideal step function (for details, see \cref{appendix: imperfect step response}). More concretely, one can write the step response as
\begin{equation}\label{definition of g(t)}
    \vartheta(t) = G_0\left[ 1 + g(t)\right]\theta(t).
\end{equation}
That is, $\vartheta(t)=0 \quad \forall t<0$, while $\vartheta(t)=G_0[1+g(t)] \quad \forall t \geq 0$. The constant factor $G_0$ can be taken care of by absorbing it into the definition of the settings to compensate the loss of the system as explained in \cref{appendix: imperfect step response}. This means that the difference between the actual step response and the ideal step function is given by $g(t)$ for $t\geq 0$. For our purposes, we are interested in systems that behave as a low-pass filter. In this case, the function $g(t)$ can be calculated as a sum of damped oscillations (see \cref{section: our transfer function} for a specific example). These oscillations are the cause of the ringing artifact that is experimentally observed. In general, this means that it is possible to find an exponential bound for $g(t)$, that is,
\begin{equation}\label{eq: bound for g}
    |g(t)| \leq Ae^{-bt} ,
\end{equation}
where the parameters $A$ and $b$  contain information about the LTI system, and thus about its bandwidth. For example, the larger the bandwidth is, the larger $b$ is expected to be, which translates into a faster response. 
As we show in \cref{appendix: imperfect step response}, when considering our phase scenario, this leads to exponential upper bounds for the correlation strengths given by
\begin{equation}\label{eq: phase correlation strength exponential}
    \overline{\epsilon}_l^{\phi}(t_0) = \frac{1}{4}A^2\Delta_{\rm max}^2e^{-2bt_0}(1+e^{-bT})^2e^{-2bT(l-1)}.
\end{equation}
Here, $\Delta_{\rm max}$ represents the worst-case scenario related to the setting choices of the $(N-l)^{\rm th}$ round. More concretely, it is the maximum difference of phase settings used in the PM for phase-encoding. In a three-state BB84 protocol, it takes the value  $\Delta_{\rm max}=\pi$. For a standard BB84 protocol using four states, however, $\Delta_{\rm max}=3\pi/2$. For further details about this parameter, see \cref{appendix: imperfect step response}. 

Similarly, if we assume that the laser pulses are so narrow (in comparison to the modulation pulses) that they can be approximated to a Dirac delta, we find exponential upper bounds for the intensity correlations given by
\begin{equation}\label{eq: intensity correlation strength exponential}
    \overline{\epsilon}_l^{\mu}(t_0) = \frac{\mu_0}{2} A \Delta_{\rm max} e^{-bt_0}\left( 1+e^{-bT} \right)e^{-b  T(l-1)}.
\end{equation}
Here, $\mu_0$ is the mean photon number corresponding to the maximum modulation intensity. In other words, it represents the mean photon number of the signal state. Again, $\Delta_{\rm max}$ represents the worst-case scenario related to the setting choices of the $(N-l)^{\rm th}$ round. Particularly, when modulating between the maximum and minimum intensity, it takes the value $\Delta_{\rm max}=\pi$. Again, for further detail we refer the reader to \cref{appendix: imperfect step response}. 

Remarkably, \cref{eq: phase correlation strength exponential,eq: intensity correlation strength exponential} are exactly of the form of \cref{eq: epsilon exponential}. Thus, we can justify the correctness of the model used in \cite{pereira2024quantum} for the model discussed above. More importantly, this allows to estimate the parameters $\overline{\epsilon}_1$ and $C$ needed to apply such security analysis. The former is calculated by simply substituting $l=1$ in \cref{eq: phase correlation strength exponential,eq: intensity correlation strength exponential}, and the latter is the factor multiplying the term $(l-1)$ in the exponents. Therefore, we find that these parameters are given by
\begin{subequations}\label{eq: definition of C eps_1 phase}
    \begin{equation}\label{eq: definition of C eps_1 phase 1}
         C^{\phi} = 2bT \quad \text{and} \quad 
    \end{equation}
    \begin{equation}\label{eq: definition of C eps_1 phase 2}
    \overline{\epsilon}_1^{\phi}(t_0) = \frac{1}{4}A^2 \Delta_{\rm max}^2 e^{-2bt_0}(1+e^{-bT})^2, 
    \end{equation}
\end{subequations}
in the case of phase correlations (indicated by the superscript) and
\begin{subequations}\label{eq: definition of C eps_1 intensity}
    \begin{equation}\label{eq: definition of C eps_1 intensity 1}
        C^{\mu} = b  T \quad \text{and} \quad 
    \end{equation}
    \begin{equation}\label{eq: definition of C eps_1 intensity 2}
        \overline{\epsilon}_1^{\mu}(t_0) = \frac{\mu_0}{2} A\Delta_{\rm max} e^{-b  t_0}\left( 1+e^{-b  T} \right), 
    \end{equation}
\end{subequations}
for intensity correlations (also indicated by the superscript).
Since $C^{\phi}$ and $C^{\mu}$ are inversely proportional to the repetition rate, this confirms the intuition that pulse correlations are stronger the faster the QKD protocol is run. However, because $b$ is related to the bandwidth of the system, it is the ratio between bandwidth and repetition rate that determines the strength of correlations and how fast they decay.

Our proposed method to characterize long-range phase and intensity correlations thus reduces to measuring the real phase waveform in order to find a bound of the form of \cref{eq: bound for g} for the function $g(t)$. Ideally, a transfer function that can reproduce the results should be found so that $A$ and $b$ may be calculated analytically. However, this is not strictly necessary, since this bound could in principle be found without knowledge of the exact oscillatory behavior of $g(t)$. A key advantage of this method is that, because of its linearity, it suffices to study only the step response of the system instead of having to generate very long sequences to estimate long-range correlations.  Moreover, note that our results for phase correlations should be applicable to any other encoding scheme using a PM. For instance, the polarization encoding in \cite{grunenfelder2020performance} is done by a PM with a birefringent crystal.  Therefore, the bit-and-basis correlations that they observed could be explained in the same way as phase correlations.

\cref{eq: definition of C eps_1 phase,eq: definition of C eps_1 intensity} allow us to calculate the effective maximum correlation length given by \cref{eq: correlation length} with parameters that can be experimentally characterized. Thus, following the method of \cite{pereira2024quantum}, we substitute $l_{\rm c}$ with $l_{\rm e}$ for any calculations that follow. The explicit expressions for $l_{\rm e}^{\phi}$ and $l_{\rm e}^{\mu}$ as well as their dependence with the number $N$ of emitted signals are shown in \cref{appendix: correlation length}.

\subsection{Experimental results}\label{section: experimental results long-range}

\begin{figure}[h!]
    \centering
    \begin{subfigure}[h]{0.45\textwidth}
        \includegraphics[width=\textwidth]{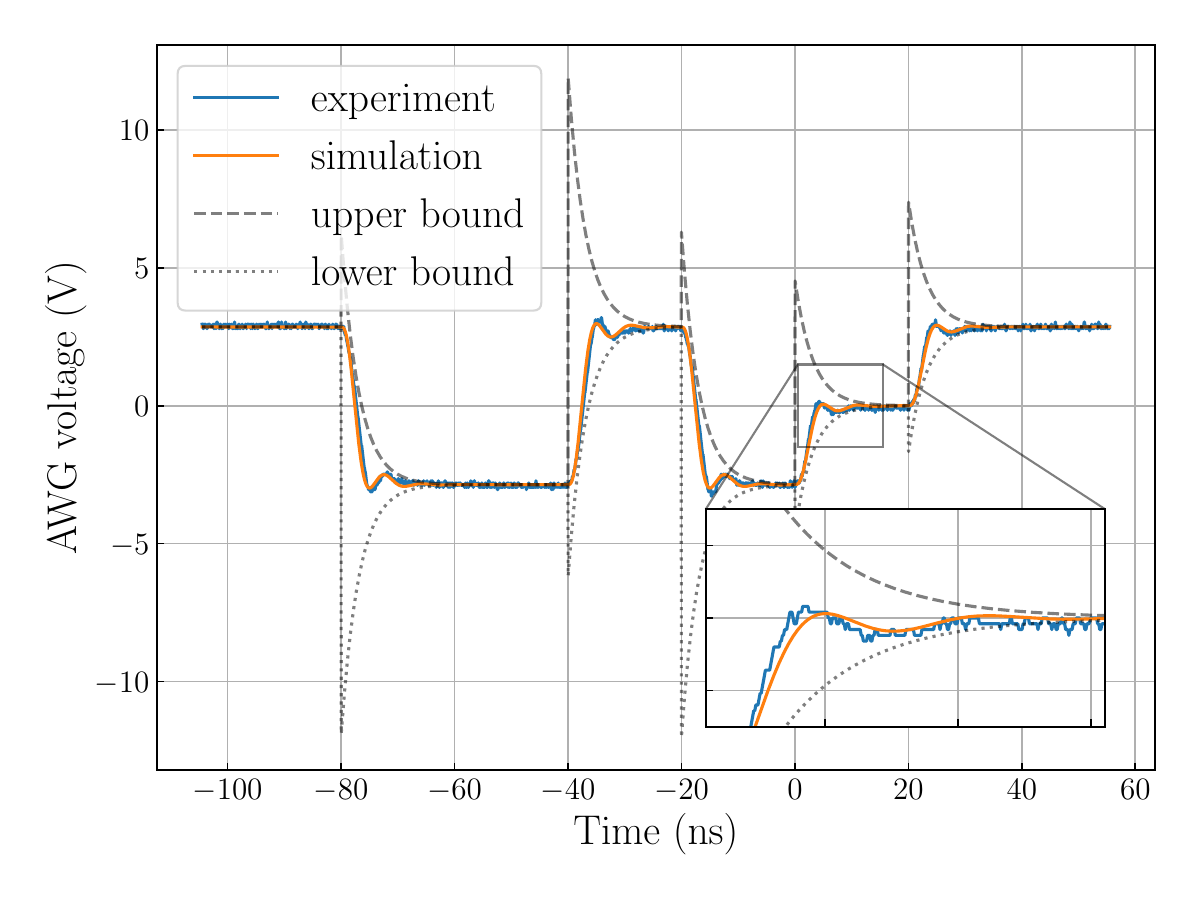}
        \caption{}
    \end{subfigure}
    \begin{subfigure}[h]{0.45\textwidth}
        \includegraphics[width=\textwidth]{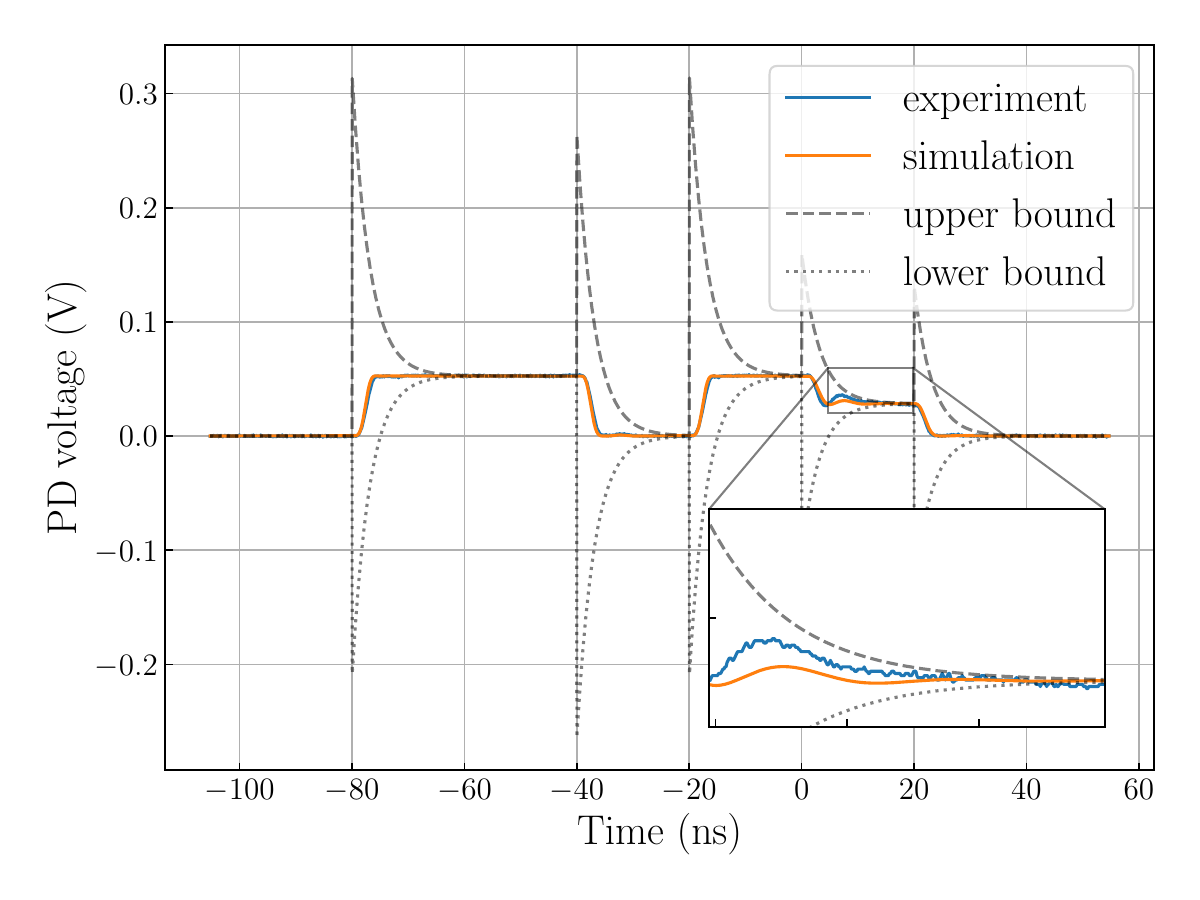}
        \caption{}
    \end{subfigure}
    \caption{(a) Voltage square pulses generated by the AWG, proportional to the phase waveform. (b) Output voltage from the PD, proportional to the intensity. The blue lines labeled as \quotes{experiment} are the voltages directly measured with the oscilloscope. The orange curves labeled as \quotes{simulation} are the results of applying the filter with fitted parameters to the perfect square, i.e. \cref{voltage in}, and the dashed lines labeled as bounds are obtained with an expression of the form of \cref{phase bound with jumps}.}
    \label{fig: fit}
\end{figure}

Conveniently, one can use the data acquired for the characterization of short-range correlations to dtermine the step response of the system. That is, no further measurements are needed to proceed with the characterization of long-range correlations.
More concretely, we use $V^{\rm AWG}(t)$ of a specific sequence to fit the model, as shown in \cref{fig: fit} (a). The presence of the ringing artifact suggests that some of the poles of the transfer function are not real. We find a transfer function made up of one real pole and two complex conjugate poles (see \cref{section: our transfer function} for more details) to provide a good approximation of the measurements. The parameters of this filter are shown in \cref{tab: filter parameters} in \cref{section: our transfer function}. Exponential bounds on the error introduced in each jump between settings are also plotted to provide an intuition of how tight they are. The parameters of the IM are also used to compare $V^{\rm PD}(t)$ with the fit in \cref{fig: fit} (b). The zoomed-in regions show parts where the model is less accurate, especially for $V^{\rm PD}(t)$, possibly due to noise from the PD. Still, the exponential bounds are such that the actual waveform stays inside, which is the only requirement to implement the proposed method. For an explicit expression of the function fitted and the bounds plotted we refer the reader to \cref{section: our transfer function}.

\subsubsection*{Estimated upper bound for phase correlations}

\begin{figure}[h]
    \centering
    \includegraphics[width=\linewidth]{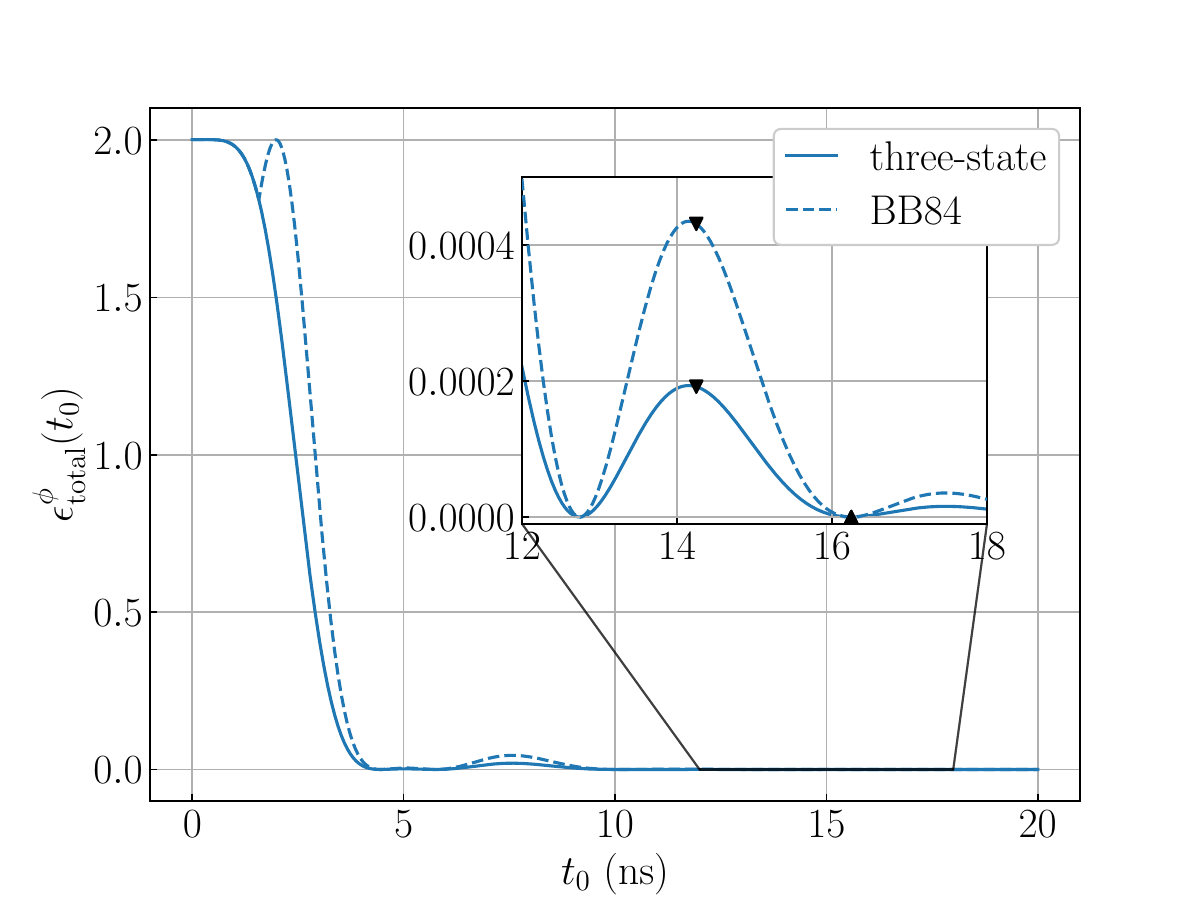}
    \caption{Plot of $\epsilon_{\rm total}^{\phi}(t_0)$ against the relative point of alignment calculated with simulated data. The dashed curve corresponds to the standard BB84 protocol ($\Delta_{\rm max}=3\pi/2$) and the solid one, to the three-state BB84 protocol ($\Delta_{\rm max}=\pi$). The triangle pointing upwards marks the minimum in the range $t_0 \in [2 \rm ns, 18 \rm ns]$, whereas the triangles pointing downwards mark the maxima inside of the 4 ns range centered around the minimum. These values are found to be $t_0^b=16.2 \rm ns$ and $t_0^w=14.2 \rm ns$, respectively.}
    \label{fig: choosing t0 for phase - sim}
\end{figure}

Using the parameters of the filter displayed in \cref{tab: filter parameters} in \cref{eq: definition of C eps_1 phase 1}, we find that $C^{\phi} \approx 12.7$. The parameter $\overline{\epsilon}_1^{\phi}(t_0)$ depends on the point of alignment. Therefore, as in \cref{section: short-range phase correlations}, we first estimate $\epsilon_{\rm total}^{\phi}(t_0)$ as a function of $t_0$ to choose two points of alignment accordingly. Since in this section we focus on the LTI model itself, we use simulated data instead of experimental data. To do so, we consider the scenario where the ideal phase signal is modified by our fitted system and use this to calculate the quantities $|\varphi_{j_N|\mathbf{\Tilde{j}}_{N-1}}(t_0)-\varphi_{j_N|\mathbf{j}_{N-1}}(t_0)|$ needed for the application of \cref{eq: experimental epsilon_l phi 2}. The results are shown in \cref{fig: choosing t0 for phase - sim}, with the chosen points $t_0^b$ and $t_0^w$ marked with triangles pointing upwards and downwards, respectively.
Note that the summations in \cref{eq:epsilon_qubit epsilon_correl} now go up to $l=l_{\rm e}^{\phi}$, which is calculated using \cref{eq: correlation length}. When considering $N=10^{12}$ emitted signals and a failure probability of $d=10^{-10}$ \cite{pereira2024quantum}, this leads to an effective maximum correlation length of $l_{\rm e}^{\phi}=6$ for both $t_0^b$ and $t_0^w$. Using the chosen points of alignment, we obtain that $\overline{\epsilon}_1^{\phi}(t_0^b) = 2.0\times 10^{-4}$ and $\overline{\epsilon}_1^{\phi}(t_0^w) = 7.2\times 10^{-4}$ for the three-state protocol ($\Delta_{\rm max}=\pi$). For the BB84 protocol ($\Delta_{\rm max}=3\pi/2$), the results are  $\overline{\epsilon}_1^{\phi}(t_0^b) = 4.5\times 10^{-4}$ and $\overline{\epsilon}_1^{\phi}(t_0^w) = 1.6\times 10^{-3}$.

\subsubsection*{Estimated upper bound for intensity correlations}\label{section: estimation of correlations}

For intensity correlations, using \cref{eq: definition of C eps_1 intensity 1} and the found parameters for the filter displayed in \cref{tab: filter parameters} in \cref{section: our transfer function}, we find that $C^{\mu} \approx 6.4$. The predicted effective first order correlation strength, however, depends on both the simulated mean photon number of a signal pulse $\mu_0$, and the point of alignment $t_0$. As in \cref{section: short-range intensity correlations}, for concreteness we choose $\mu_0=0.3$. For the points of alignment, we take again the ones obtained in \cref{section: short-range intensity correlations}. This yields $\overline{\epsilon}_1^{\mu}(t_0^b) = 3.8\times 10^{-3}$ and $\overline{\epsilon}_1^{\mu}(t_0^w) = 5.3 \times 10^{-3}$. If we consider $N=10^{12}$ emitted signals and  a failure probability of $d=10^{-10}$, this yields an effective maximum correlation length of $l_{\rm e}^{\mu}=11$ for both values of  $t_0$ considered.

\section{Secret-key rate simulations}\label{section: SKR simulations}

We use the security proof of \cite{LTQCoin} combined with the analysis of \cite{pereira2024quantum} to calculate the finite secret-key rate (SKR) for a standard BB84 protocol, with phase-encoded single photons, in the presence of phase correlations of arbitrary length. Note that we do not consider intensity correlations because the current security analyses handling this imperfection \cite{Zapatero_2021,Sixto_2022} are in the asymptotic regime. Therefore, we cannot directly apply them to the finite key scenario nor employ the result in \cite{pereira2024quantum} to include the case of unbounded intensity correlations. Also, we remark that for the simulations we consider the BB84 protocol just as an example. Since the correlation strengths for the BB84 and three-state protocols are comparable, the results are expected to be similar.

The security proof of \cite{LTQCoin} requires the estimation of the parameter $\epsilon_{\rm total}^{\phi}(t_0)$ defined in \cref{eq:epsilon_qubit epsilon_correl}, but using $l_{\rm e}^\phi$ to extend it to correlations of unbounded length following the methodology in \cite{pereira2024quantum}. For further details, see \cref{appendix: calculation of epsilon}. The repetition rate is set to the one of our experiment, 50MHz. The number of emitted pulses is $N=10^{12}$. The assumed error correction efficiency is $f=1.16$, the dark count probability of Bob's detectors is $p_{d}=10^{-8}$, and their detector efficiency is assumed to be ideal. The channel transmission rate is $\eta = 10^{-0.02 L}$, where $L$ is the channel distance in kilometers. The correctness and secrecy parameters are set to $\epsilon_{\rm correctness}=10^{-15}$ and $\epsilon_{\rm secrecy}=10^{-9}$, respectively.

\begin{figure}[h]
    \centering
    \includegraphics[width=\linewidth]{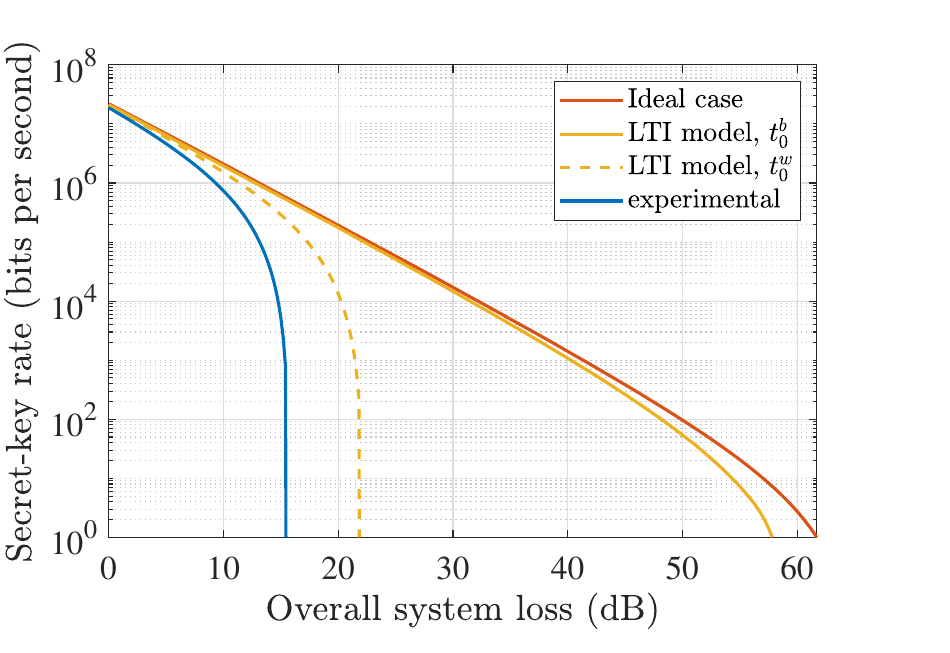}
    \caption{Secret-key rates calculated using the security proof of \cite{LTQCoin}. The ideal case (orange line) shows the achievable SKR with perfect state preparation. The blue and yellow lines take unbounded bit-and-basis correlations into account with the analysis of \cite{pereira2024quantum}. For both sets of lines, the long-range correlations are accounted for using the exponential bounds $\overline{\epsilon}_l^{\phi}(t_0)$ given by \cref{eq: phase correlation strength exponential}. The blue line shows the achievable rate that is estimated with the best-case experimental values of the short-range correlation strengths. The yellow lines correspond to the achievable secret key rates when $\epsilon_l^{\phi}(t_0)$ are estimated with the simulated phase modulation pulses and \cref{eq: experimental epsilon_l phi 2}. The solid and dashed lines are calculated, respectively, for the points $t_0^b$ and $t_0^w$ marked with triangles in \cref{fig: choosing t0 for phase - sim}.} 
    \label{fig:SKR results}
\end{figure}

In \cref{fig:SKR results}, we show the results for a few different scenarios. Firstly, we plot for reference the ideal case (orange line), in which there are no correlations nor SPFs (i.e. $\epsilon_{\rm total}^{\phi}(t_0)=0$ $\forall  t_0$). In the other lines, we consider bit-and-basis correlations by calculating $\epsilon_{\rm total}^{\phi}(t_0)$ in different ways. For all of them, we set the failure probability to $d=10^{-10}$ \cite{pereira2024quantum} and the resulting estimated effective maximum correlation length is $l_{\rm e}^{\phi}=6$. In blue, we plot the SKR estimated with the best-case experimental correlation strengths plotted in \cref{fig: results phase correlations}. These values correspond to $l=1,2,3,4$. For higher orders, the exponential bounds given by \cref{eq: phase correlation strength exponential} are used. This yields $\epsilon_{\rm total}^{\phi}(t_0^b)=1.94\times 10^{-3}$. Lastly, in yellow we plot the results obtained when using a hybrid method in which the exponential bounds are assumed for long-range correlations, while for short-range we use the correlation strengths that are estimated from analytical results given by the LTI model. More concretely, we directly calculate $\varphi_{j_N|\mathbf{\Tilde{j}}_{N-1}}(t_0)-\varphi_{j_N|\mathbf{j}_{N-1}}(t_0)$ with \cref{phase deviation} derived in \cref{appendix: imperfect step response} to apply it to \cref{eq: experimental epsilon_l phi 2}. The solid and dashed curves correspond to using the points of alignment $t_0^{b}$ and $t_0^{w}$ shown in \cref{fig: choosing t0 for phase - sim}, respectively. In this case, the obtained values for $\epsilon^{\phi}(t_0)$ are $3.83\times 10^{-10}$ and $4.31\times 10^{-4}$, respectively. 

The performance when considering the experimental values is lower than when directly using the LTI model. This is so because the estimated experimental correlations strengths can include other imperfections which are not purely SCD correlations. Since SPFs, contrary to correlations, have a very low impact on the SKR \cite{LTQCoin}, this is an important distinction to make. On the other hand, the high drop in performance between the solid yellow line (best-case) and the dashed yellow line (worst-case) hints that when positioning ourselves into a bad point of alignment, we approach the pessimistic exponential bounds for short-range correlations. This highlights the importance of choosing correctly a point of alignment for better performance. And vice-versa, this also reveals a risk that could arise if one is not careful with the alignment. If, by chance, one falls into a good point of alignment when characterizing correlations, one could estimate them to be very low. However, if one is not aware that this could be highly dependent on the alignment, the actual protocol could have much higher correlations. Thus, we could be underestimating the amount of information that is being leaked. On the other hand, in a real experiment, such a precise alignment might not be feasible. Thus, it could be beneficial for the performance of the protocol to simply lower the repetition rate. This model could also be viewed in this way as a tool to gauge whether the best strategy is to push for high repetition rates or keep them well below the bandwidth of the system to considerably reduce the effect of correlations.

\section{Conclusions }\label{section: conclusions}

Setting-choice dependent pulse correlations caused by bandwidth limitations are difficult to characterize experimentally when no underlying model is assumed. This is so because the number of different sequences that need to be considered grows exponentially with the maximum correlation length. This issue becomes even more apparent when we consider the fact that correlations may, in general, be of unbounded length, which means it would be necessary to generate an arbitrarily large number of possible combinations of settings. For this reason, the method proposed in this work, which models the response of the system, is a very practical way to account for correlations in a QKD setup. It relies on a linear assumption, which agrees well with the experimental results observed. 

We have applied the proposed method to an experimental setup imitating the phase and intensity modulation inside a transmitter for QKD. In this setup, the main restriction is the bandwidth of the AWG (150MHz), but we remark that our model is device agnostic and can be adapted to other setups where the main limitation might be, for example, an optical modulator. This is so because it works as a black box that takes an ideal phase signal and applies a transfer function that could in principle accommodate the imperfection of several devices.

The linear model provides, to the best of our knowledge, the first ever estimation of long-range correlations going up to an infinite order. Moreover, the found upper bounds on the strength of correlations decrease exponentially with the pulse separation, fitting the security proof introduced in \cite{pereira2024quantum}. The rate at which they decrease is inversely proportional to the repetition rate at which the QKD protocol is run. At the same time, it is proportional to a parameter that is related to the bandwidth of the system. This confirms that the strength of correlations depend on the ratio between bandwidth and repetition rate. This knowledge can also help the user decide if it is more beneficial for the performance of the QKD protocol to use a very high repetition rate and account for the correlations that this provokes, or lower the repetition rate to reduce the impact of this side-channel.

The secret-key rate results presented in \cref{section: SKR simulations} emphasize that, on the one hand, distinguishing between correlations and SPFs is crucial for performance. This is made clear in \cref{fig:SKR results}, where even the line calculated with the best-case experimental values is lower than the worst-case of the LTI model because of the effect of experimental noise. On the other hand, they also highlight the importance of careful alignment between modulation and signal pulses, both for achieving good performance and for preventing an underestimation of the actual information leakage during the key exchange.

\begin{acknowledgments}

We thank Daniil Trefilov and Xoel Sixto for valuable discussions. We acknowledge support from the Galician Regional Government (consolidation of Research Units: AtlantTIC), the Spanish Ministry of Economy and Competitiveness (MINECO), the Fondo Europeo de Desarrollo Regional (FEDER) through the grant No. PID2020-118178RB-C21, MICIN with funding from the European Union NextGenerationEU (PRTR-C17.I1) and the Galician Regional Government with own funding through the \quotes{Planes Complementarios de I+D+I con las Comunidades Autonomas} in Quantum Communication, the \quotes{Hub Nacional de Excelencia en Comunicaciones Cuánticas} funded by the Spanish Ministry for Digital Transformation and the Public Service and the European Union NextGenerationEU, the European Union's Horizon Europe Framework Programme under the Marie Sklodowska-Curie Grant No. 101072637 (Project QSI) and the project \quotes{Quantum Security Networks Partnership} (QSNP, grant agreement No 101114043) and the European Union via the European Health and Digital Executive Agency (HADEA) under the Project 101135225 — QuTechSpace.

\end{acknowledgments}

\section*{Author contributions}
D.R. conceptualized the work. A.A. and F.G. conducted the experiment. A.A. developed the main theoretical results, with inputs from D.R., M.P., G.C.-L and M.C. M.P. and G.C.-L. provided the code for the SKR simulations. D.R. and H.Z. supervised the work. A.A. wrote the manuscript with inputs from all of the authors towards improving it and checking the validity of the results.

\onecolumngrid
\appendix

\newpage
\section{Estimation of intensity correlations from AWG}\label{Appendix: intensity correlations from AWG}

\begin{figure}[h!]
    \centering
    \begin{subfigure}[h]{0.45\textwidth}
        \includegraphics[width=\linewidth]{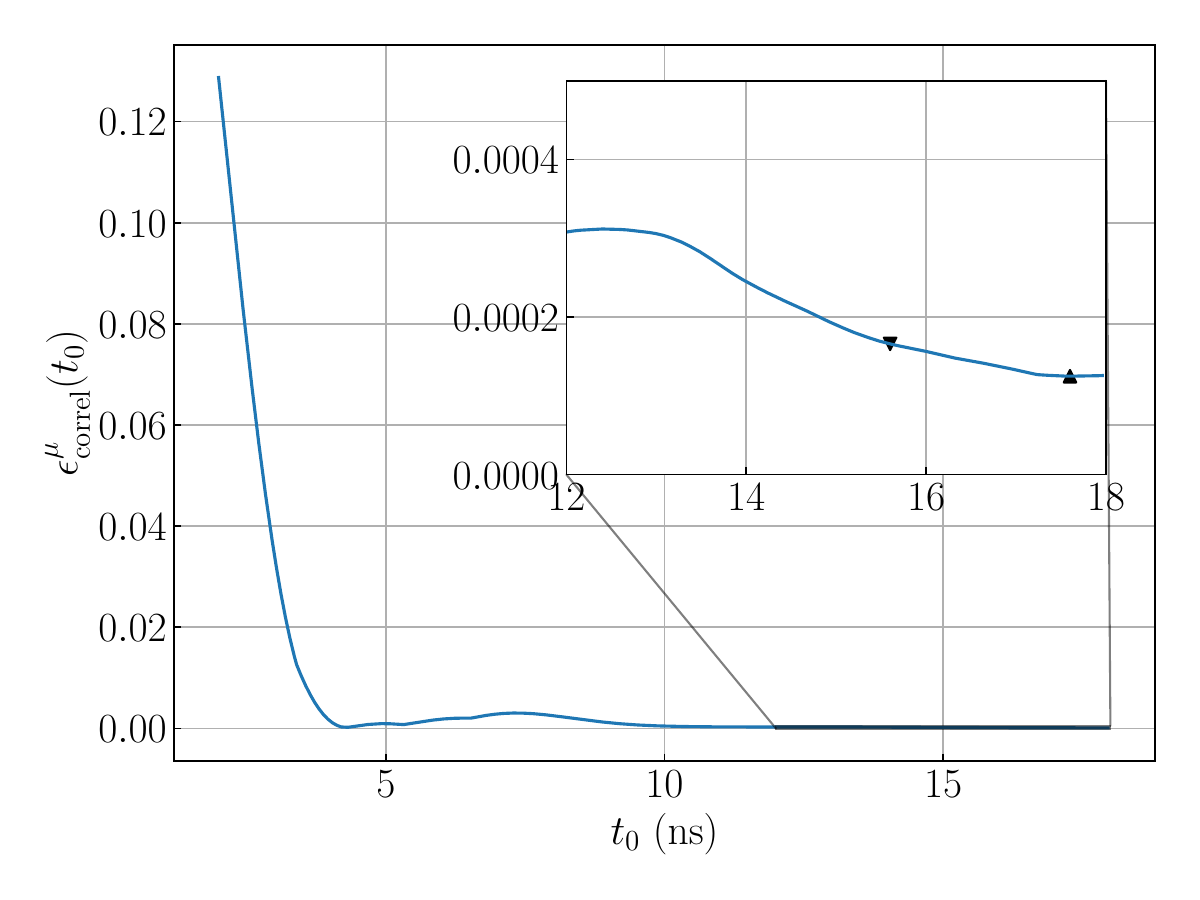}
        \caption{}
    \end{subfigure}
    \begin{subfigure}[h]{0.45\textwidth}
        \includegraphics[width=\linewidth]{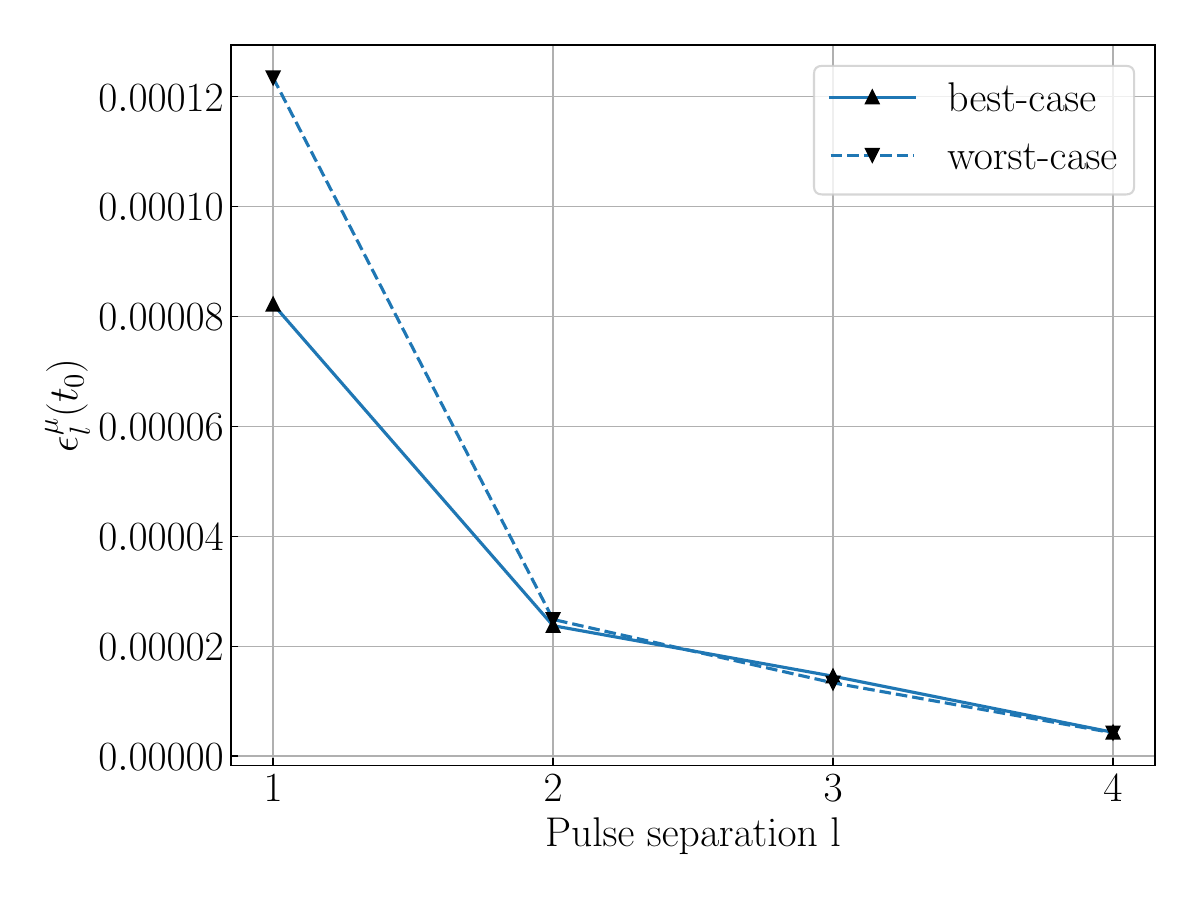}
        \caption{}
    \end{subfigure}
    \caption{(a) $\epsilon_{\rm correl}^{\mu}(t_0)$ against the relative point of alignment inside the modulation pulse calculated with data from the AWG. The chosen points of alignment are $t_0^{b}=17.6 \rm ns$ and $t_0^{w}=15.6 \rm ns$.
    (b)  Maximum intensity correlation strength estimated for $t_0^{b}$ and $t_0^{w}$ against the distance between signals $l=1,...,4$.}
    \label{fig: deciding t0 intensity experimental appendix}
\end{figure}

In this Appendix, we show how intensity correlations can be experimentally characterized from $V^{\rm AWG}(t)$ as mentioned in \cref{section: short-range intensity correlations}.
When considering intensity correlations, we make the assumption that they are solely caused by imperfections in the phase signal controlling the PM within the MZI. We consider that the PM introduces an imperfect phase $\varphi(t)$ so that the intensity after interference $I_{out}(t)$ is given by 
\begin{equation}\label{eq: MZI}
    I_{out}(t) = \frac{I_{in}(t)}{2}\left( 1 + \sin\varphi(t) \right),
\end{equation}
where $I_{in}(t)$ is the intensity of the light that is injected into the MZI. Consider now a laser pulse. Its mean photon number after going through the IM is proportional to the energy of the modulated laser pulse, that is
\begin{equation}\label{eq: mu phi}
    \mu(t_0) \propto \int_{t_0-\Delta t/2}^{t_0 +\Delta t /2} 
    f(t-t_0)(1+\sin\varphi(t))dt.
\end{equation}
Like in \cref{section: short-range intensity correlations}, we assume that $f(t-t_0)$ is given by \cref{eq: gaussian pulse}, although the same calculations could be done for any other shape. \cref{eq: mu phi} implies that memory effects in $\varphi(t)$ will result in intensity correlations, which can then be estimated through \cref{eq: experimental intensity correlations}. Thus, instead of using \cref{eq: mu out}, one can estimate $\mu$ directly from the AWG with the following expression:
\begin{equation}\label{eq: mu Vin}
    \mu(t_0) \propto \int_{t_0-\Delta t/2}^{t_0 +\Delta t /2} 
    f(t-t_0)\left(1+\sin\frac{V^{\rm AWG}(t)}{V_{\pi}}\pi\right) dt,
\end{equation}
where we have used \cref{eq: action of PM} to relate $\varphi(t)$ with $V^{\rm AWG}(t)$. The results of applying this method are shown in \cref{fig: deciding t0 intensity experimental appendix} for $t_0^{b}=17.6 \rm ns$ and $t_0^{w}=15.6 \rm ns$.

\newpage
\section{Calculation of $\epsilon^{\phi}$ for the SKR simulations }\label{appendix: calculation of epsilon}

In this Appendix, we show how to use the experimental knowledge acquired to estimate the achievable SKR of a QKD protocol whose emitted pulses are correlated across rounds. For concreteness, we consider the case of bit-and-basis pulse correlations induced by an imperfect PM, which has been addressed by the security proofs in \cite{RTpereira_quantum_2020,mizutaniFinitekeySecurity2023,pereira2023,LTQCoin}. As explained in the main text, these analyses assume that such correlations have a finite and known maximum length $l_{\rm c}$, i.e., that a particular pulse is only influenced by the previous $l_{\rm c}$ setting choices. However, using the formalism in \cite{pereira2024quantum}, these proofs can be extended to a more realistic scenario in which the length of these correlations is unbounded. 

As discussed in the main text, this entails considering an effective maximum correlation length $l_{\rm e}^{\phi}$ such that the residual correlations are negligible. If the correlation strength $\epsilon^{\phi}_l$ decays exponentially with the pulse separation $l$, as in \cref{eq: epsilon exponential}, then the effective maximum correlation length $l_{\rm e}^{\phi}$ takes the form of \cref{eq: correlation length}. In what follows, we explain how to relate $\epsilon^{\phi}_l$ and $l_{\rm e}^{\phi}$ with the parameters needed to apply the security proof in \cite{LTQCoin}.

This security proof requires an upper bound on the parameter $\epsilon^{\phi}$, which is defined as \cite{LTQCoin}
\begin{equation}
    \epsilon^{\phi} \leq \epsilon^{\phi}_{\rm qubit} + \epsilon^{\phi}_{\rm correl},
    \label{eq:epsilon}
\end{equation}
where $\epsilon^{\phi}_{\rm qubit}$ quantifies how much the $k^{\rm th}$ pulse differs from a certain qubit state, given its dependence on the previous $l_{\rm e}^{\phi}$ setting choices; and $\epsilon^{\phi}_{\rm correl}$ quantifies the amount of information about the $k^{\rm th}$ setting choice $j_k$ that is leaked through the following $l_{\rm e}^{\phi}$ pulses. By combining the results in \cite{LTQCoin} and \cite{pereira2024quantum}, one can upper bound these quantities by 
\begin{equation}
    \epsilon^{\phi}_{\rm qubit} \leq 1 - \big|\braket{\phi_{j_k}|\psi_{j_k|j_{k-1},j_{k-2},\hdots,j_{k-l_{\rm e}^{\phi}},\gamma,\hdots,\gamma}}\big|^2 ~~~\text{and}~~~ \epsilon^{\phi}_{\rm correl} \leq \sum_{l=1}^{l_{\rm e}^{\phi}} \epsilon^{\phi}_l.
    \label{eq:equbit_ecorrel}
\end{equation}
In \cref{eq:equbit_ecorrel}, $\ket{\psi_{j_k|j_{k-1},j_{k-2},\hdots,j_{k-l_{\rm e}^{\phi}},\gamma,\hdots,\gamma}}$ is the state emitted in the $k^{\rm th}$ round conditioned on the previous setting choices being $j_{k-1},j_{k-2},\hdots,j_{k-l_{\rm e}^{\phi}},\gamma,...,\gamma$, where $\gamma,...,\gamma$ is a fixed sequence of setting choices from the $(k-l_{\rm e}^{\phi}-1)^{\rm th}$ to the $1^{\rm st}$ round, e.g., $\gamma=0_Z$; and $\{\ket{\phi_{j_k}}\}_{j_k}$ is any set of qubit states that only depend on the $k^{\rm th}$ setting choice $j_k$. Since $\epsilon^{\phi}_{\rm correl}$ is already written as a function of $\epsilon^{\phi}_l$ and $l_{\rm e}^{\phi}$, we only need to derive an expression for $\epsilon^{\phi}_{\rm qubit}$. 

For this, we start by defining $\ket{\phi_{j_k}} \coloneqq \ket{\psi_{j_k|\gamma,\gamma,...,\gamma}}$, where $\ket{\psi_{j_k|\gamma,\gamma,...,\gamma}}$ is the state emitted on the $k^{\rm th}$ round conditioned on a fixed sequence for all previous setting choices. Note that this state has the form of \cref{phase-encoded qubit} for some phase values $\varphi_{j_k|\gamma,\gamma,...,\gamma}$, and therefore $\{\ket{\psi_{j_k|\gamma,\gamma,...,\gamma}}\}_{j_k}$ is a set of qubit states, as required. Then, using the relationship between the fidelity and the trace distance we have that
\begin{align}
&\big|\braket{\psi_{j_k|\gamma,\gamma,...,\gamma}|\psi_{j_k|j_{k-1},j_{k-2},...,j_{k-l_{\rm e}^{\phi}},\gamma,...,\gamma}}\big| = \sqrt{1 - T\left(\ket{\psi_{j_k|\gamma,\gamma,...,\gamma}},\ket{\psi_{j_k|j_{k-1},j_{k-2},...,j_{k-l_{\rm e}^{\phi}},\gamma,...,\gamma}}\right)^2},
\label{eq:fid_tra}
\end{align}
where $T(\ket\cdot,\ket\cdot)$ denotes the trace distance between two pure states. Next, following a similar approach to that in \cite{pereira2024quantum}, we use the triangle inequality consecutively to bound the trace distance term in \cref{eq:fid_tra} as 
\begin{align}
&T\left(\ket{\psi_{j_k|\gamma,\gamma,...,\gamma}},\ket{\psi_{j_k|j_{k-1},j_{k-2},...,j_{k-l_{\rm e}^{\phi}},\gamma,\hdots,\gamma}}\right) \nonumber \\
    &\leq T\left(\ket{\psi_{j_k|\gamma,\gamma,...,\gamma}},\ket{\psi_{j_k|j_{k-1},\gamma,...,\gamma}}\right) + T\left(\ket{\psi_{j_k|j_{k-1},\gamma,...,\gamma}},\ket{\psi_{j_k|j_{k-1},j_{k-2},...,j_{k-l_{\rm e}^{\phi}},\gamma,\hdots,\gamma}}\right) \nonumber \\
    &\leq T\left(\ket{\psi_{j_k|\gamma,\gamma,...,\gamma}},\ket{\psi_{j_k|j_{k-1},\gamma,...,\gamma}}\right) + T\left(\ket{\psi_{j_k|j_{k-1},\gamma,...,\gamma}},\ket{\psi_{j_k|j_{k-1},j_{k-2},\gamma,...,\gamma}}\right) \nonumber \\
    &+ T\left(\ket{\psi_{j_k|j_{k-1},j_{k-2},\gamma,...,\gamma}},\ket{\psi_{j_k|j_{k-1},j_{k-2},j_{k-3},\gamma,...,\gamma}}\right) \nonumber \\
&+...+T\left(\ket{\psi_{j_k|j_{k-1},...,j_{k-l_{\rm e}^{\phi}+1},\gamma,\hdots,\gamma}},\ket{\psi_{j_k|j_{k-1},...,j_{k-l_{\rm e}^{\phi}},\gamma,\hdots,\gamma}}\right).
    \label{eq:tri_inq}
\end{align}
Using again the relationship between the fidelity and the trace distance, we have that \cref{eq:tri_inq} can be expressed as 
\begin{align}
&T\left(\ket{\psi_{j_k|\gamma,\gamma,...,\gamma}},\ket{\psi_{j_k|j_{k-1},j_{k-2},...,j_{k-l_{\rm e}^{\phi}},\gamma,\hdots,\gamma}}\right) \nonumber \\
&\leq \sqrt{1-\left|\braket{\psi_{j_k|\gamma,\gamma,...,\gamma}|\psi_{j_k|j_{k-1},\gamma,...,\gamma}}\right|^2} + ... + \sqrt{1-\left|\braket{\psi_{j_k|j_{k-1},...,j_{k-l_{\rm e}^{\phi}+1},\gamma,\hdots,\gamma}|\psi_{j_k|j_{k-1},...,j_{k-l_{\rm e}^{\phi}},\gamma,\hdots,\gamma}}\right|^2}.
\label{eq:tra_fid}
\end{align}
Then, using the fact that the strength of the correlation $\epsilon^{\phi}_l$ between pulses separated by $l$ rounds can be quantified by
\begin{align}
&\big|\braket{\psi_{j_k|j_{k-1},...,j_{k-l+1},\tilde{j}_{k-l},j_{k-l-1},...,j_1}|\psi_{j_k|j_{k-1},...,j_{k-l+1},{j}_{k-l},j_{k-l-1},...,j_1}}\big|^2 \geq 1 - \epsilon^{\phi}_{l},
\label{eq:assumption}
\end{align}
where $\tilde{j}_{k-l}$ is a setting choice that may differ from $j_{k-l}$,  \cref{eq:tra_fid} becomes
\begin{equation}
T\left(\ket{\psi_{j_k|\gamma,\gamma,...,\gamma}},\ket{\psi_{j_k|j_{k-1},j_{k-2},...,j_{k-l_{\rm e}^{\phi}},\gamma,...,\gamma}}\right) \leq \sqrt{\epsilon^{\phi}_1} + ... + \sqrt{\epsilon^{\phi}_{l_{\rm e}^{\phi}}} = \sum_{l=1}^{l_{\rm e}^{\phi}} \sqrt{\epsilon^{\phi}_l}.
\label{eq:tra_bound}
\end{equation}
Substituting \cref{eq:tra_bound} in \cref{eq:fid_tra}, and then in \cref{eq:equbit_ecorrel}, we find that 
\begin{equation}
\label{eq:bound_equbit}
    \epsilon^{\phi}_{\rm qubit}  = \left(\sum_{l=1}^{l_{\rm e}^{\phi}} \sqrt{\epsilon^{\phi}_l}\right)^2. 
\end{equation}
Finally, using \cref{eq:epsilon}, we can express $\epsilon^{\phi}$ as a function of $\epsilon^{\phi}_l$ and $l_{\rm e}^{\phi}$ as
\begin{equation}
\epsilon^{\phi} \leq \left(\sum_{l=1}^{l_{\rm e}^{\phi}} \sqrt{\epsilon^{\phi}_l}\right)^2 + \sum_{l=1}^{l_{\rm e}^{\phi}} \epsilon^{\phi}_l.
\end{equation}
Having obtained a bound on $\epsilon^{\phi}$, one can directly apply the security analysis in \cite{LTQCoin} to estimate the secret-key length, and consequently the SKR for different values of $N$. As an example, we consider the BB84 protocol and simulate the achievable SKR as a function of the overall system loss. The results are presented in \cref{fig:SKR results}. {For simplicity, in these simulations, we have assumed that 
$\{\ket{\psi_{j_k|\gamma,\gamma,...,\gamma}}\}_{j_k}$ are ideal BB84 states. In practice, these may deviate from the ideal BB84 states due to SPFs, however, these are known to have a very minor impact on the SKR \cite{LTQCoin}}.

\newpage
\section{Dependence of intensity correlations with chosen setting}\label{appendix: setting dependence}

\begin{figure}[h!]
    \centering
    \begin{subfigure}[h]{0.45\textwidth}
        \includegraphics[width=\textwidth]{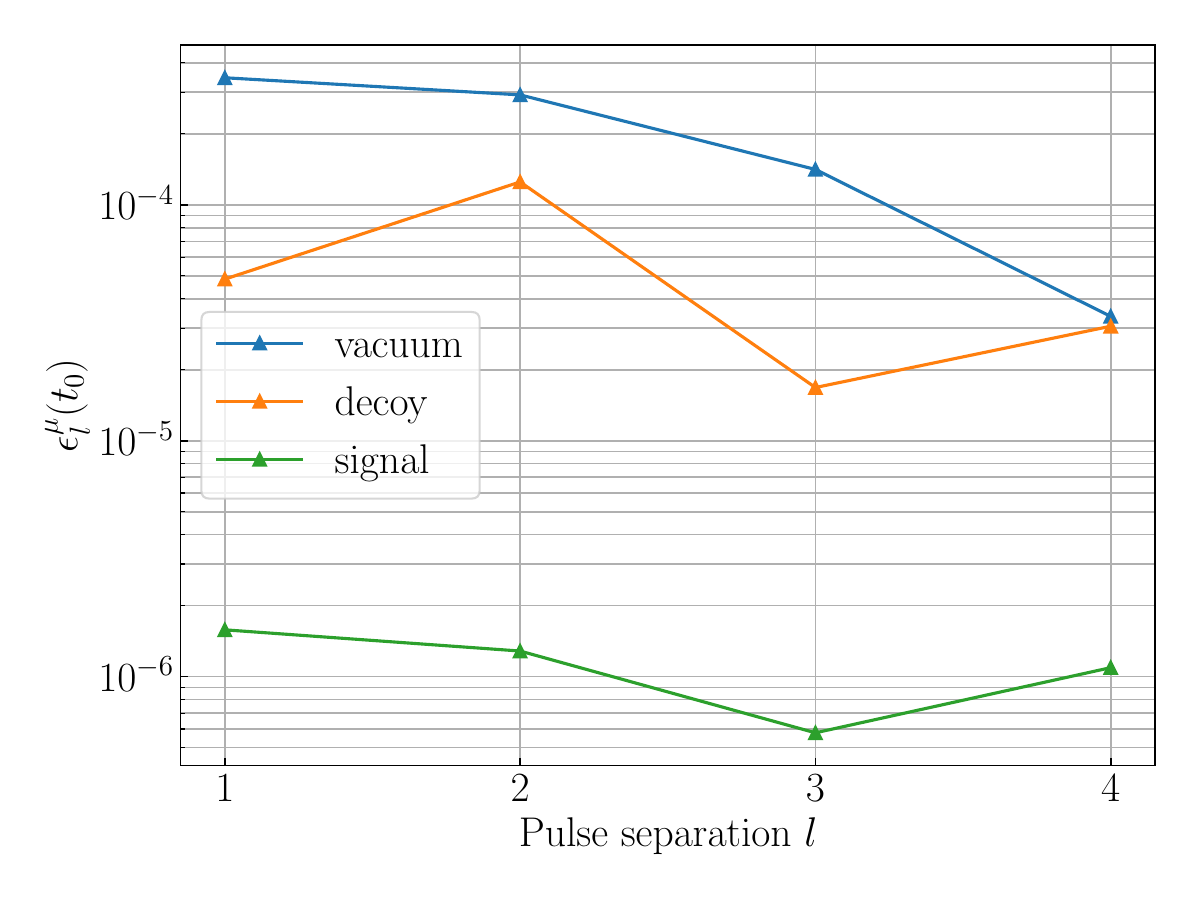}
    \end{subfigure}
    \caption{Intensity correlation strengths at $t_0^{b}=16.6 \rm ns$ plotted against the distance between pulses $l=1,...,4$. In blue, data corresponding to the vacuum level. In orange, data corresponding to the decoy level. In green, data corresponding to the signal level.}
    \label{fig: intensity correlations PD according to settings}
\end{figure}

In this Appendix, we show that the strength of intensity correlations varies greatly with the setting of the emitted pulse. In \cref{fig: intensity correlations PD according to settings}, the intensity correlation strengths are plotted for each state (i.e., fixing the setting $j_N$ to either vacuum, signal or decoy and maximizing the quantity defined in \cref{eq: experimental intensity correlations} over the past setting choices only). There is a clear difference between the observed correlation strengths when changing the last setting choice, which does not happen with phase correlations. 

There are two causes for this difference. On the one hand, because the action of an MZI is not linear with the phase (see \cref{eq: MZI}), the same error in phase results in a different error in intensity depending on the phase value. More concretely, for small perturbations of the phase $\delta \varphi$, we have that $\delta I \sim \cos(\varphi)\delta \varphi$. Hence, the effect will be much more strongly felt around the maximum slope point (decoy) than at the extrema (vacuum and signal). On the other hand, contrary to \cref{eq: experimental epsilon_l phi 2},  which only depends on the phase error, \cref{eq: experimental intensity correlations} also depends on the actual intensity $\mu$ value of the state. Moreover, from \cref{eq: fidelity of two Poissonians II} it follows that, for two intensities $\mu$ and $\mu + \delta \mu $ such that $\mu>0$ and $\delta \mu\geq 0$, their fidelity satisfies the inequality $F_{\mu, \mu + \delta \mu}\geq F_{0, \delta \mu}$, which explains why the correlation strengths of the vacuum are the highest.

\newpage
\section{How correlations originate from an imperfect step response}\label{appendix: imperfect step response}

In this Appendix, we explicitly show how SCD correlations arise naturally when considering a transfer function for the system and how to arrive to the exponential upper bounds for the correlations given by \cref{eq: phase correlation strength exponential} and \cref{eq: intensity correlation strength exponential}.

Let $H(s)$ be the transfer function of an LTI system, where $s$ is a complex variable $s=\sigma + i \omega$ whose imaginary part $\omega$ is the angular frequency. $H(s)$ characterizes the behavior of the system according to the frequency of the input.  The inverse Laplace transform of $H(s)$, $h(t) = \mathcal{L}^{-1}\left[H(s)\right](t)$, is known as the impulse response. Let us consider that the system is given an input $\phi(t)$. The output of the system, which we denote by $\varphi(t)$, can be calculated in the s-domain as $H(s)\mathcal{L}[\phi(t)](s)$. In the time domain, this is equivalent to calculating the convolution of the impulse response and $\phi(t)$, that is,
\begin{equation}
    \varphi (t) 
    = (h * \phi)(t)
    = \int^t_0 d\tau   h(\tau)  \phi(t-\tau).
\end{equation}
When the considered input is the Heaviside or step function, which we denote by $\theta(t)$ and is defined as $\theta(t)=1$ $\forall t\geq 0$ and $\theta(t) = 0$ otherwise, the result is the so-called step response, given by
\begin{equation}
    \vartheta(t) 
    = (h * \theta)(t)
    =\int _0^t d\tau   h(\tau).
\end{equation}

In a QKD experiment using different phase or intensity levels, one needs to generate a phase signal made up of square pulses whose heights are chosen from a set of fixed values. These values are the so-called settings of the experiment, which we denote by $j_k$. This means that the input of the LTI system is given by
\begin{equation}\label{eq: perfect square pulses}
    \phi(t) = \sum_{k=1}^N j_k \left[ \theta(t -(k-1)T) - \theta(t-kT) \right].
\end{equation}
Here, $T$ is the inverse of the repetition rate, but also the width of the pulses for simplicity, as this is the case in many experimental implementations. The linearity of the system implies that the output $\varphi(t)$ can be calculated with the step response $\vartheta(t)$ as
\begin{equation}\label{output}
    \varphi(t) = \sum_{k=1}^N j_k \left[ \vartheta(t -(k-1)T) - \vartheta(t-kT) \right].
\end{equation}
Without loss of generality, $\vartheta(t)$ can be rewritten as
\begin{equation}
    \vartheta(t) = G_0\left[ 1 + g(t)\right]\theta(t), \tag{\ref{definition of g(t)}}
\end{equation}
where the function $g(t)$ may be regarded as the difference between the step response and the step function when $G_0=1$. The parameter $G_0$ is a constant factor, and to simplify the formulas below it is absorbed in the definition of the settings as $j'_k := G_0 j_k$. In practice this entails changing the settings in order to compensate for the loss. We find, therefore, that \cref{output} may be rewritten as
\begin{equation}\label{phi_out}
    \varphi(t) = \phi'(t) + \sum_{k=1}^N j'_k \left[  g(t-(k-1)T) \theta(t-(k-1)T) - g(t-kT) \theta(t-kT) \right] ,
\end{equation}
where $\phi'(t)$ is the ideal signal given by \cref{eq: perfect square pulses} after changing the coefficients $j_k$ to $j'_k$. Note that the summation introduces two imperfections in the signal. On the one hand, each time bin has now a dependence on the setting choices of the previous rounds. On the other hand,  in general, the signal is no longer constant inside each time-bin. Thus, the summation in \cref{phi_out} explains the classical SCD pulse correlations due to memory effects and the side-channel due to a time-varying encoding \cite{gnanapandithan_hidden_2025}. Here, we focus only on the former.

In order to characterize correlations of a certain order $l$, we consider two sequences of phase modulation pulses with the same choice of settings except for the $(N-l)^{\rm th}$ round. Following the same notation as in \cref{section: theoretical framework},  we denote them by $\varphi_{j_N|\mathbf{j}_{N-1}}(t)$ and $\varphi_{j_N|\mathbf{\Tilde{j}}_{N-1}}(t)$ . Then, their difference is given by
\begin{equation}\label{phase deviation}
    \varphi_{j_N|\mathbf{\Tilde{j}}_{N-1}}(t_0)-\varphi_{j_N|\mathbf{j}_{N-1}}(t_0) = 
    \Delta_{N-l}
    \Bigl[g(t_0+lT) - g\left(t_0 + (l-1)T\right)\Bigr],
\end{equation}
where we drop the step function because we consider $(N-1)T \leq t \leq NT$, i.e., we focus on the $N^{\rm th}$ pulse. The parameter $t_0$ is defined as $t_0:= t-(N-1)T$ with $0\leq t_0 \leq T$. It represents the point inside the modulation pulse in which the states are positioned. Lastly, $\Delta_{N-l}$ is defined as $\Delta_{N-l} := \Tilde{j}'_{N-l} - j'_{N-l}$, i.e., it is the difference of the setting chosen in the $(N-l)^{\rm th}$ round between the sequences in consideration.

As stated in \cref{section: exponential upper bounds for correlations}, we are interested in systems that behave as a low-pass filter. Consider, for instance, a transfer function with $n$ non-degenerate complex poles $s_k=\omega_k e^{i(\pi-\alpha_k)}$ of the form

\begin{equation}
    H(s) =  G_0\prod_{k=1}^{n}\frac{\omega_k}{s-s_k}.
\end{equation}
Its step response can be shown to be given by
\begin{equation}
    \vartheta(t) =  G_0 \sum_{k=1}^n c_k \left( e^{s_k t} - 1 \right) \theta(t),
\end{equation}
where $c_k$ are some complex coefficients that depend on the chosen poles. To obtain the correct behavior, we choose the poles such that their real part is negative (this ensures an exponential decay, so that the steady state can be reached). Furthermore, for every non-real pole, its complex conjugate should also be added, which ensures that $\vartheta(t)$ is real. Meeting these requirements, the error function $g(t)$ defined in \cref{definition of g(t)} can be expected to be a sum where each pole contributes an oscillatory term. The angular frequency of each of the oscillations is equal to the imaginary part of the corresponding pole. Each term is weighted by some real coefficient and an exponential factor with the exponent being the real part of the corresponding pole. This means that, as we show with a concrete example in \cref{section: our transfer function}, it is possible to find a bound for $g(t)$ of the form of \cref{eq: bound for g}, which states that
\begin{equation}
    |g(t)| \leq Ae^{-bt} . \tag{\ref{eq: bound for g}}
\end{equation}
Here, the parameters $A$ and $b$ hold information about the poles and, therefore, about the bandwidth of the system. Using \cref{eq: bound for g} in \cref{phase deviation} we have that
\begin{equation}\label{exponential phase bound}
    |\varphi_{j_N|\mathbf{\Tilde{j}}_{N-1}}(t_0)-\varphi_{j_N|\mathbf{j}_{N-1}}(t_0)| \leq
    \Delta_{\rm max}\Bigl[|g(t_0+lT)| + |g\left(t_0 + (l-1)T|\right)\Bigr] \leq
    A \Delta_{\rm max} e^{-bt_0}\left( 1+e^{-bT} \right)e^{-b  T(l-1)},
\end{equation}
where $\Delta_{\rm max}$ is the maximum difference between the settings $j'_{N-l}$ and $\Tilde{j}'_{N-l}$.

In the following subsections, we apply the result given by \cref{exponential phase bound} to bound both phase and intensity correlations.

\subsection*{Upper-bounding phase correlations}

Consider two qubits of the form of \cref{phase-encoded qubit} with phases $\varphi_{j_N|\mathbf{j}_{N-1}}$ and $\varphi_{j_N|\mathbf{\Tilde{j}}_{N-1}}$ as defined in \cref{section: theoretical framework}. Their fidelity is given by
\begin{equation}\label{eq:explicit fidelity calculation  (phase)}
    \left|\braket{\psi(\varphi_{j_N|\mathbf{\Tilde{j}}_{N-1}})|\psi(\varphi_{j_N|\mathbf{j}_{N-1}})}\right|^2 =
    \frac{1}{4}\left|  \left(1+e^{i(\varphi_{j_N|\mathbf{j}_{N-1}} - \varphi_{j_N|\mathbf{\Tilde{j}}_{N-1}})}\right) \right|^2 =
    \cos^2\left(\frac{\varphi_{j_N|\mathbf{j}_{N-1}} - \varphi_{j_N|\mathbf{\Tilde{j}}_{N-1}}}{2}\right),
\end{equation}
where in the last equality we have used the trigonometric identity $\frac{1}{2}(1+\cos 2x)=
    \cos^2x$. Thus, the fidelity can be lower bounded by using the fact that $\cos x \geq 1-x^2/2$, so that
\begin{equation}
\begin{split}
    &\left|\braket{\psi(\varphi_{j_N|\mathbf{\Tilde{j}}_{N-1}})|\psi(\varphi_{j_N|\mathbf{j}_{N-1}})}\right|^2 
    \geq 
    \left[ 1 - \frac{1}{2}\left(\frac{\varphi_{j_N|\mathbf{\Tilde{j}}_{N-1}} - \varphi_{j_N|\mathbf{j}_{N-1}}}{2}\right)^2  \right]^2
    \\ &= 1  - \left(\frac{\varphi_{j_N|\mathbf{\Tilde{j}}_{N-1}} - \varphi_{j_N|\mathbf{j}_{N-1}}}{2}\right)^2  + \frac{1}{4}\left(\frac{\varphi_{j_N|\mathbf{\Tilde{j}}_{N-1}} - \varphi_{j_N|\mathbf{j}_{N-1}}}{2}\right)^4 \geq 1 - \left(\frac{\varphi_{j_N|\mathbf{\Tilde{j}}_{N-1}} - \varphi_{j_N|\mathbf{j}_{N-1}}}{2}\right)^2.
\end{split}
\end{equation}
Since this is already of the form given by \cref{definition epsilon_l}, we find that 
\begin{equation}
    \overline{\epsilon}_l^{\phi}(t_0)
    =\max_{\mathbf{j}_N,\mathbf{\Tilde{j}}_{N}} \left\{
    \left(\frac{\varphi_{j_N|\mathbf{\Tilde{j}}_{N-1}}(t_0) - \varphi_{j_N|\mathbf{j}_{N-1}}(t_0)}{2}\right)^2 
    \right\}
    \leq 
    \frac{1}{4}A^2\Delta_{\rm max}^2e^{-2bt_0}(1+e^{-bT})^2e^{-2bT(l-1)},
\end{equation} 
where in the last inequality we use \cref{exponential phase bound}. Thus, we recover the upper bound given by \cref{eq: phase correlation strength exponential}.

\subsection*{Upper-bounding intensity correlations}

In this section, we make the approximation that the laser pulses are so narrow compared to the modulation pulses that one can approximate them to a Dirac delta, so that \cref{eq: mu phi} becomes
\begin{equation}\label{action of MZI on PRWCPs (approximation)}
    \mu(t_0) = \frac{\mu_0}{2}\left( 1 + \sin\varphi(t_0) \right),
\end{equation}
where $\mu_0$ denotes the mean photon number of the input pulses to the MZI. 

Consider two PR-WCPs with mean photon numbers $\mu_{j_N|\mathbf{j}_{N-1}}$ and $\mu_{j_N|\mathbf{\Tilde{j}}_{N-1}}$ previously defined in \cref{section: theoretical framework}. These values come from modulating the input pulses with mean photon number $\mu_0$ by using relative phases $\varphi_{j_N|\mathbf{j}_{N-1}}$ and $\varphi_{j_N|\mathbf{\Tilde{j}}_{N-1}}$ in a balanced MZI, respectively. Without loss of generality, let $\mu_{j_N|\mathbf{\Tilde{j}}_{N-1}} - \mu_{j_N|\mathbf{j}_{N-1}}\geq 0$. Then, 
\begin{equation}
\begin{split}
    &\mu_{j_N|\mathbf{\Tilde{j}}_{N-1}} - \mu_{j_N|\mathbf{j}_{N-1}} 
    = \frac{\mu_0}{2}\left|\sin\varphi_{j_N|\mathbf{\Tilde{j}}_{N-1}}-\sin\varphi_{j_N|\mathbf{j}_{N-1}}\right| 
    \\& = \mu_0 \left|\sin\left(\frac{\varphi_{j_N|\mathbf{\Tilde{j}}_{N-1}} - \varphi_{j_N|\mathbf{j}_{N-1}}}{2}\right) \cos\left( \frac{\varphi_{j_N|\mathbf{\Tilde{j}}_{N-1}} + \varphi_{j_N|\mathbf{j}_{N-1}}}{2}  \right) \right| 
    \leq \mu_0 \left| \sin\left(\frac{\varphi_{j_N|\mathbf{\Tilde{j}}_{N-1}} - \varphi_{j_N|\mathbf{j}_{N-1}}}{2}\right) \right| .
\end{split}
\end{equation}
Since $|\sin (x)| \leq |x|$, we obtain that
\begin{equation}\label{bound for delta_mu}
    \mu_{j_N|\mathbf{\Tilde{j}}_{N-1}} - \mu_{j_N|\mathbf{j}_{N-1}} \leq \frac{\mu_0}{2}|\varphi_{j_N|\mathbf{\Tilde{j}}_{N-1}} - \varphi_{j_N|\mathbf{j}_{N-1}}|.
\end{equation}

This means that we can use the upper bound on the phase difference given by \cref{exponential phase bound} to upper bound the mean photon number difference. 

The fidelity of the two PR-WCPs is given by
\begin{equation}\label{fidelity of two Poissonians}
\begin{split}
    &F_{\mu_{j_N|\mathbf{j}_{N-1}}, \mu_{j_N|\mathbf{\Tilde{j}}_{N-1}} } = 
    \left( \sum_{n=0}^{\infty} \sqrt{p(n|\mu_{j_N|\mathbf{j}_{N-1}})p(n|\mu_{j_N|\mathbf{\Tilde{j}}_{N-1}})} \right)^2 
    \\& = \left(  \sum_{n=0}^{\infty} \frac{\sqrt{e^{-(\mu_{j_N|\mathbf{j}_{N-1}}+\mu_{j_N|\mathbf{\Tilde{j}}_{N-1}})}(\mu_{j_N|\mathbf{j}_{N-1}}\mu_{j_N|\mathbf{\Tilde{j}}_{N-1}})^n}}{n!} \right)^2,
\end{split}
\end{equation}
where we have used that, for a Poissonian distribution,
\begin{equation}\label{eq: Poissonian distribution}
    p(n|\gamma) = \frac{e^{-\gamma}\gamma^n}{n!} ,
\end{equation}
with $\gamma \in \{\mu_{j_N|\mathbf{j}_{N-1}}, \mu_{j_N|\mathbf{\Tilde{j}}_{N-1}}\}$. Rewriting \cref{fidelity of two Poissonians} in terms of $\mu_{j_N|\mathbf{\Tilde{j}}_{N-1}}- \mu_{j_N|\mathbf{j}_{N-1}}$, we find that
\begin{equation}\label{eq: fidelity of two Poissonians II}
    F_{\mu_{j_N|\mathbf{j}_{N-1}}, \mu_{j_N|\mathbf{\Tilde{j}}_{N-1}} } = e^{-(\mu_{j_N|\mathbf{\Tilde{j}}_{N-1}}- \mu_{j_N|\mathbf{j}_{N-1}})}
    \left[
    \sum_{n=0}^{\infty} p(n|\mu_{j_N|\mathbf{j}_{N-1}})\left( 1+\frac{\mu_{j_N|\mathbf{\Tilde{j}}_{N-1}}- \mu_{j_N|\mathbf{j}_{N-1}}}{\mu_{j_N|\mathbf{j}_{N-1}}} \right)^{n/2}
    \right]^2.
\end{equation}
Since $\mu_{j_N|\mathbf{\Tilde{j}}_{N-1}} - \mu_{j_N|\mathbf{j}_{N-1}}\geq 0$, the sum is bigger than unity and, therefore, we have that
\begin{equation}
\begin{split}
    &F_{\mu_{j_N|\mathbf{j}_{N-1}},  \mu_{j_N|\mathbf{\Tilde{j}}_{N-1}}} 
    \geq e^{-(\mu_{j_N|\mathbf{\Tilde{j}}_{N-1}}- \mu_{j_N|\mathbf{j}_{N-1}})}
    \geq 1 - (\mu_{j_N|\mathbf{\Tilde{j}}_{N-1}}- \mu_{j_N|\mathbf{j}_{N-1}})
    \\ & \geq 1 - \frac{\mu_0}{2}|\varphi_{j_N|\mathbf{\Tilde{j}}_{N-1}}- \varphi_{j_N|\mathbf{j}_{N-1}}| 
    \geq
    1 - \frac{\mu_0}{2} A \Delta_{\rm max} e^{-bt_0}\left( 1+e^{-bT} \right)e^{-b  T(l-1)},
\end{split}
\end{equation}
where in the second inequality we have used the fact that $e^{-x} \geq 1-x$ for real $x$, in the third inequality we have applied \cref{bound for delta_mu} and in the fourth inequality we have used \cref{exponential phase bound}.

\newpage
\section{Transfer function used for the fit}\label{section: our transfer function}
In this Appendix, we show the transfer function and step response of the LTI system that is used for the fit of the experimental data. We also show how to find the parameters $A$ and $b$ from \cref{eq: bound for g}. Lastly, we give details about how the fit is done.

\begin{figure}[h]
    \centering
    \includegraphics[width=0.4\linewidth]{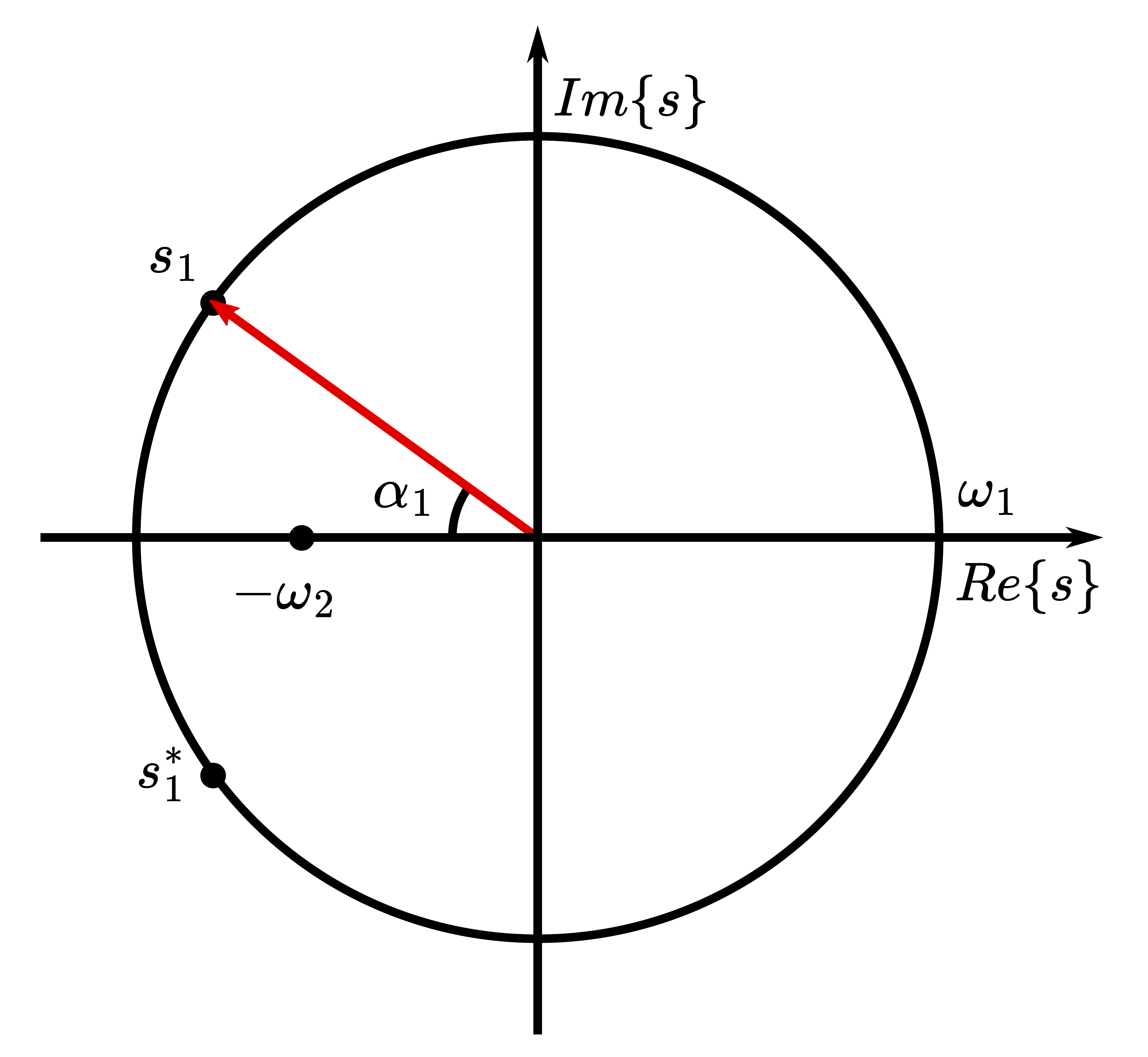}
    \caption{Plot of the pole $s_1=\omega_1e^{i(\pi - \alpha_1)}$, its complex conjugate and $s_2=-\omega_2$ in the complex plane.}
    \label{fig: three poles}
\end{figure}

We consider a transfer function given by a complex pole $s_1 = \omega_1 e^{i (\pi-\alpha_1)}$, its complex conjugate $s_1^{*}$ and a real pole $s_2=-\omega_2$ as shown in \cref{fig: three poles}, that is,
\begin{equation}\label{three poles}
    H(s) = G_0 \frac{\omega_1^2 \omega_2}{(s-s_1)(s-s_1^{*})(s+\omega_2)}.
\end{equation}
The resulting step response is given by
\begin{equation}\label{step response three poles}
    \vartheta(t) = G_0 \left[
    1 - \frac{\omega_1\omega_2}{\sin\alpha_1}\left(
    \frac{\sin\eta}{\omega_2  r} e^{-\omega_2  t}
    + \frac{1}{\omega_1  r} e^{-\omega_1 \cos\alpha_1   t}\sin\left( \omega_1  \sin\alpha_1  t + \alpha_1 - \eta \right)
    \right)\right] \theta(t),
\end{equation}
where $r$ and $\eta$ are the modulus and phase of $s_1+\omega_2$, respectively, that is,
\begin{equation}
    r := \sqrt{\omega_1^2 + \omega_2^2 - 2\omega_1\omega_2\cos\alpha_1 } \quad , \quad \eta := \arctan\frac{\omega_1\sin\alpha_1}{\omega_2 - \omega_1\cos\alpha_1}.
\end{equation}
From equation \cref{step response three poles}, it is easy to see that the error function as defined in \cref{definition of g(t)} may be written as
\begin{equation}
    g(t) = -\frac{\omega_1\omega_2}{\sin\alpha_1}\left(
    \frac{\sin{\eta}}{\omega_2  r}e^{-\omega_2  t} + 
    \frac{1}{\omega_1  r}e^{-\omega_1  \cos\alpha_1  t}\sin\left( \omega_1  \sin\alpha_1  t + \alpha_1 - \eta \right)
    \right).
\end{equation}
This can be easily bounded for $t\geq 0$ with
\begin{equation}\label{g bound for three poles}
    |g(t)| \leq \frac{\omega_1\omega_2}{\sin\alpha_1}\left(
    \frac{\sin{\eta}}{\omega_2  r} + 
    \frac{1}{\omega_1  r}
    \right)e^{-\min\{\omega_2, \omega_1  \cos\alpha_1\} t},
\end{equation}
so that
\begin{equation}\label{A and c for three poles}
    A=\frac{\omega_1\omega_2}{\sin\alpha_1}\left(
    \frac{\sin{\eta}}{\omega_2  r} + 
    \frac{1}{\omega_1  r}
    \right) ~~ \text{and} ~~ b=\min\{\omega_2, \omega_1  \cos\alpha_1\}.
\end{equation}

\subsection*{Fit of the experimental data}

In order to find the filter that best resembles our setup, we do a least squares fit of the data of $V^{\rm AWG}(t)$ to
\begin{equation}\label{voltage in}
\begin{split}
    V^{\rm AWG}(t) = & \sum_{k=1}^N j_k \left[ \vartheta(t_k; \nu_1, \nu_2, \alpha_1) - \vartheta(t_{k+1};  \nu_1, \nu_2, \alpha_1) \right]\\
    &+j_0(1-\vartheta(t_0; \nu_1, \nu_2, \alpha_1)) + j_0\vartheta(t_{N+1}; \nu_1, \nu_2, \alpha_1) ,
\end{split}
\end{equation}
where $j_0$ is the setting of the steady state, in this case, $V_{\pi}/2$. The variable $t_k$ is defined as $t_k \equiv t - (k-1)T$. The function $\vartheta(t; \nu_1, \nu_2, \alpha_1)$ is the step response given by \cref{step response three poles} using the poles $s_1=2\pi \nu_1e^{j(\pi-\alpha_1)}$, $s_1^*=2\pi \nu_1e^{-j(\pi-\alpha_1)}$ and $s_2=-2\pi \nu_2$. The results of this fit are shown in \cref{fig: fit} and the optimal parameters are displayed in \cref{tab: filter parameters}. The upper and lower bounds that are also shown in \cref{fig: fit} come from rearranging \cref{phi_out} in terms of the \quotes{jumps} between consecutive settings as
\begin{equation}
    \varphi(t) - \phi'(t) = (j'_1 - j'_0) g(t_1)\theta(t_1)  + \sum_{k=2}^N (j'_{k}-j'_{k-1}) g(t_k)\theta(t_k) + 
    (j'_0-j'_N) g(t_{N+1}) \theta(t_{N+1}),
\end{equation}
and then bounding its absolute value by
\begin{equation}\label{phase bound with jumps}
\begin{split}
    |\varphi(t) - \phi'(t)| 
    &\leq
    |j'_1 - j'_0| |g(t_1)|\theta(t_1)  + \sum_{k=2}^N |j'_{k}-j'_{k-1}| |g(t_k)|\theta(t_k) + 
    |j'_0-j'_N| |g(t_{N+1})| \theta(t_{N+1}) \\
    &\leq 
    |j'_1 - j'_0|A e^{-b  t_1}\theta(t_1)  + \sum_{k=2}^N |j'_{k}-j'_{k-1}|A e^{-c  t_k}\theta(t_k) + 
    |j'_0-j'_N|A e^{-b  t_{N+1}}\theta(t_{N+1}).
\end{split}
\end{equation}

\begin{table}[h]
\centering
\begin{tabular}{cccccc}
    \hline
   $G_0$  & $\nu_1$ (MHz) & $\nu_2$ (MHz) & $\alpha_1$ (rad) & $A$ & $b$ (MHz) \\
   \hline
   0.95  & 164 & 80 & 1.26 &  1.60 & 318.7 \\
   \hline
\end{tabular}
\caption{Optimal filter parameters for the experimental results and parameters of the bound for the error function $g(t)$.}
\label{tab: filter parameters}
\end{table}

\newpage
\section{Effective maximum correlation lengths}\label{appendix: correlation length}

In this Appendix, we show the explicit expressions for the effective maximum correlation lengths that can be derived from our model. More concretely, by taking the parameters defined in \cref{eq: definition of C eps_1 phase 1,eq: definition of C eps_1 phase 2,eq: definition of C eps_1 intensity 1,eq: definition of C eps_1 intensity 2} and plugging them into \cref{eq: correlation length}, we have that
\begin{equation}\label{effective correlation length (phase)}
    l_{\rm e}^{\phi}(N, d, t_0) =\left\lceil
    \frac{1}{2b  T} \ln\left[ \frac{ N A^2 \Delta_{\rm max}^2 \left(  1+e^{-b  T} \right)^2}{4d^2\left( 
    1-e^{-b  T} \right)^2} \right] - \frac{t_0}{T}
    \right\rceil
\end{equation}
and 
\begin{equation}\label{effective correlation length (intensity)}
    l_{\rm e}^{\mu}(N, d, t_0) = \left\lceil
    \frac{1}{b  T} \ln\left[ \frac{N\mu_0 A \Delta_{\rm max} \left(  1+e^{-b  T} \right)}{2d^2\left( 
    1-e^{-b  T/2} \right)^2} \right] - \frac{t_0}{T}
    \right\rceil,
\end{equation}
where $\lceil x\rceil$ denotes the ceiling of $x$. We take the ceiling because correlation lengths must be integer numbers.

\begin{figure}[h!]
    \centering
    \begin{subfigure}[h]{0.45\textwidth}
        \includegraphics[width=\textwidth]{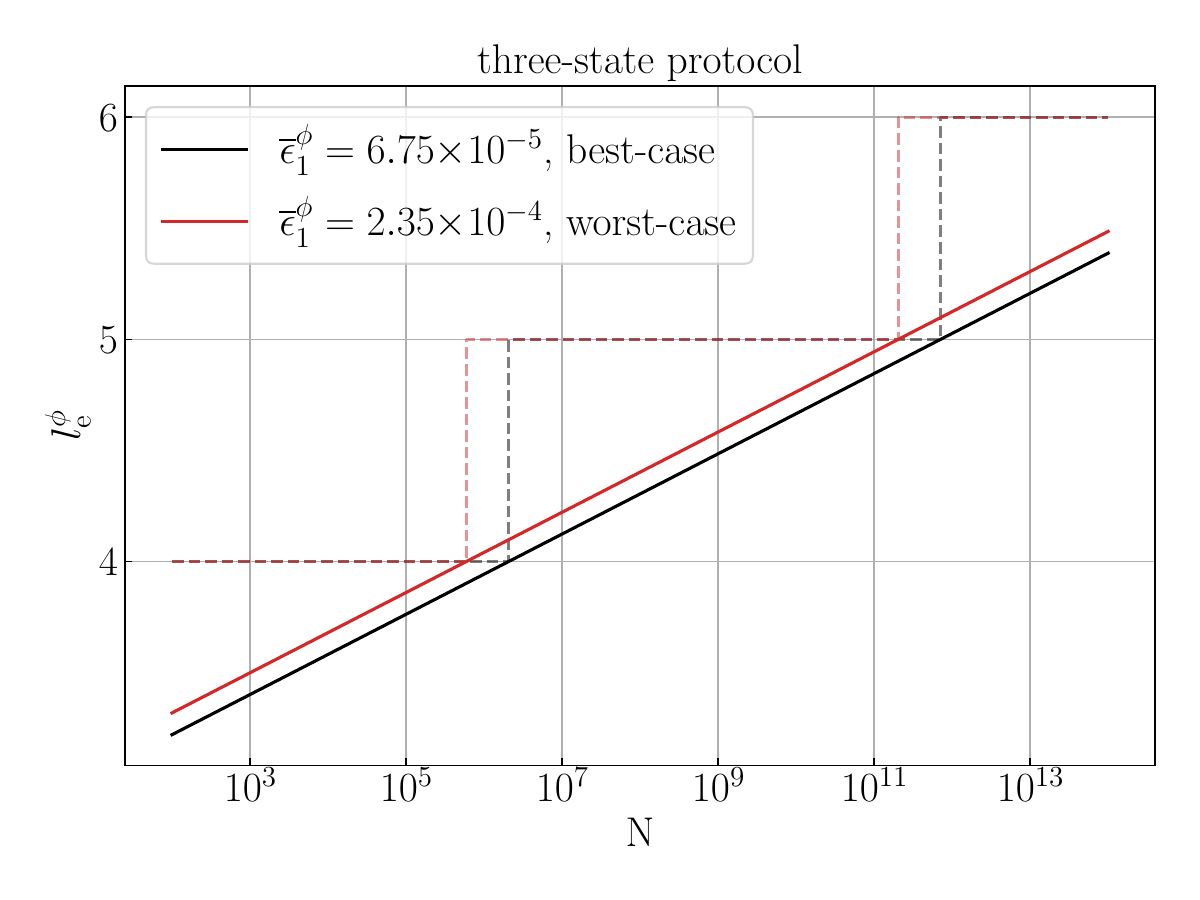}
        \caption{}
    \end{subfigure}
    \begin{subfigure}[h]{0.45\textwidth}
        \includegraphics[width=\textwidth]{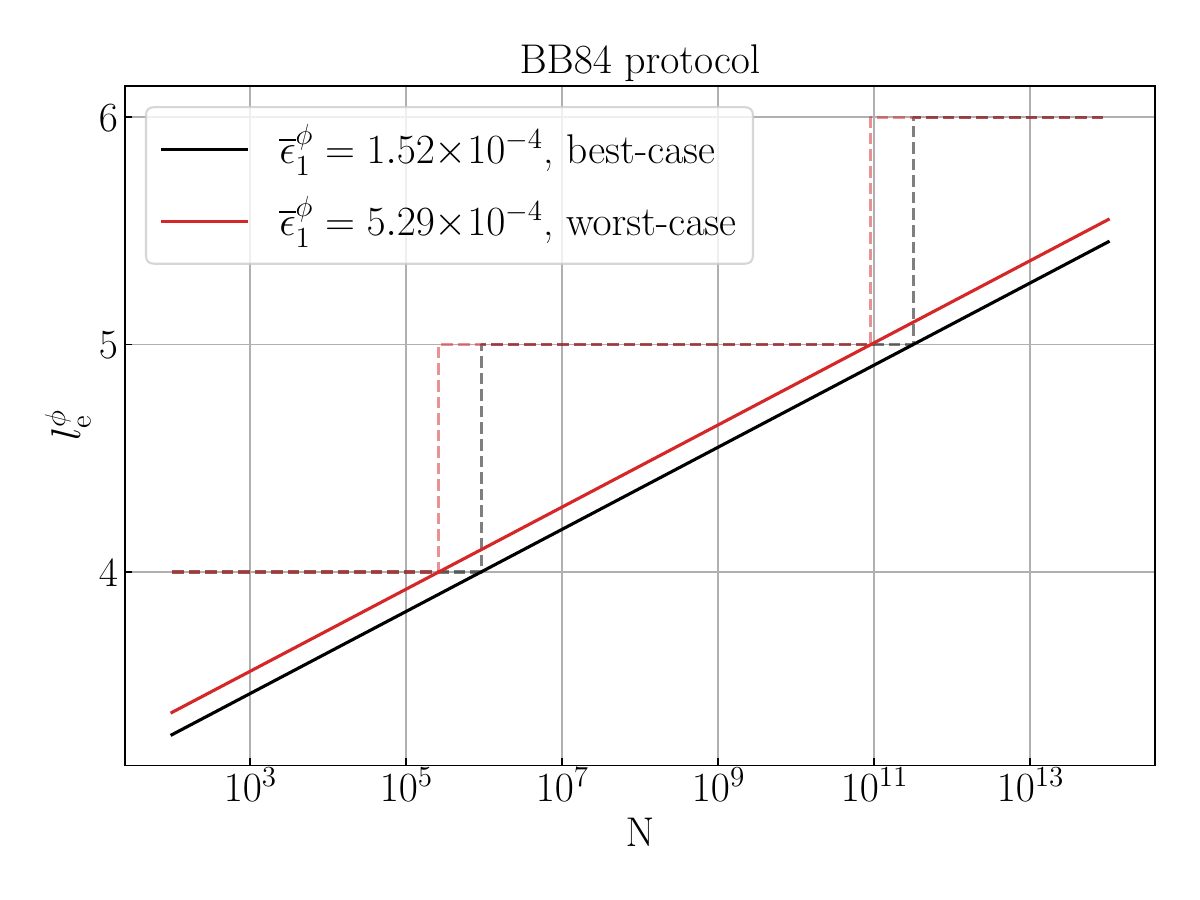}
        \caption{}
    \end{subfigure}
    \caption{(a) Effective maximum correlation length as a function of the number of emitted pulses in a phase-encoding scenario for a three-state protocol. The black(red) curves assume the best(worst)-case scenario, which, following the same notation as in the main text, refers to the point of alignment $t_0^w$($t_0^b$). The dashed lines are the ceiling of the solid lines. (b) Effective maximum correlation length as a function of the number of emitted pulses in a phase-encoding scenario for a standard BB84 protocol. The black(red) curves assume the best(worst)-case scenario, that is, the point of alignment $t_0^b$($t_0^w$). The dashed lines are the ceiling of the solid lines.}
    \label{fig:l_e (phase)}
\end{figure}

\begin{figure}[h]
    \centering
    \begin{subfigure}[h]{0.45\linewidth}
        \includegraphics[width=\textwidth]{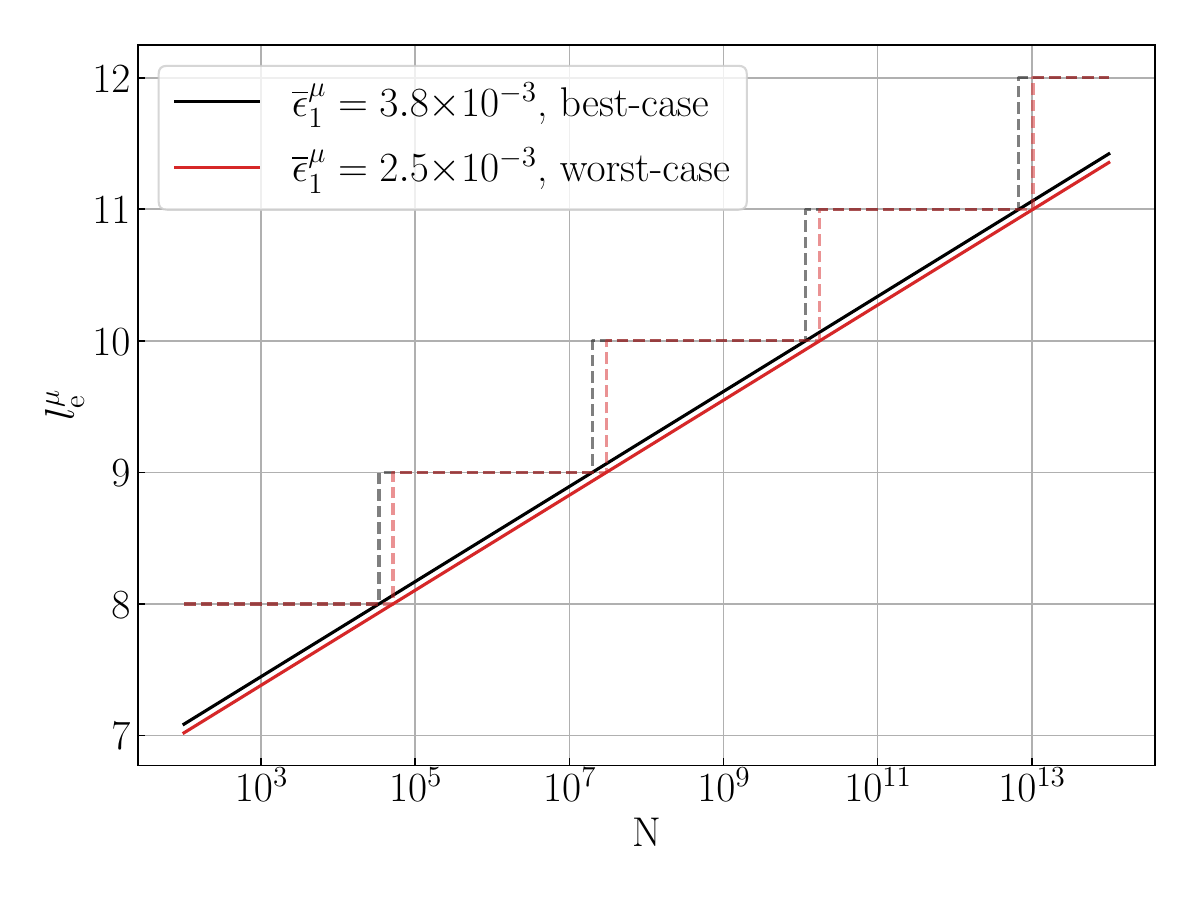}
    \end{subfigure}
    \caption{Effective maximum correlation length for the scenario considering intensity correlations. The black(red) curves assume the best(worst)-case scenario, i.e. the point of alignment $t_0^b$ ($t_0^w$). The dashed lines are the ceiling of the solid lines.}
    \label{fig: effective maximum intensity correlation length results}
\end{figure}

Additionally, in \cref{fig:l_e (phase),fig: effective maximum intensity correlation length results} we plot the effective maximum correlation lengths as a function of the number of emitted pulses $N$.

\newpage
\bibliography{bibliography.bib}

\end{document}